\documentclass[11pt]{article}
\RequirePackage{amsmath, amssymb, mathtools, amsthm, mathrsfs}
\RequirePackage{xcolor, graphicx, multirow, array, booktabs}
\RequirePackage[framemethod=tikz]{mdframed} 
\RequirePackage[colorlinks, linkcolor=blue, citecolor=blue, urlcolor=blue, bookmarks, bookmarksnumbered, pagebackref=false]{hyperref}
\RequirePackage{xfrac} 
\RequirePackage[inline]{enumitem} 
\RequirePackage{cite} 
\RequirePackage{tikz} 
\usetikzlibrary{arrows.meta, arrows, fadings}
 \tikzfading[name=fade out, inner color=transparent!0, 
             outer color=transparent!100]
 \def\corner#1#2#3#4#5%
     {\begin{scope}
       \clip #1 -- ++({#2}:{#4*1.5})
                arc ({#2}:{#3}:{#4*1.5}) -- cycle;
       \filldraw[#5, path fading=fade out] #1 circle (#4);
      \end{scope}}

\theoremstyle{plain}

 \newtheorem{cor}{Corollary}[section]
\theoremstyle{definition}
 
 \newtheorem{rem}{Remark}[section]
 
\numberwithin{equation}{section}
\def\rd{\mathrm{d}}
\def\ri{\mathrm{i}\mkern1mu}
\def\IC{\mathbb{C}}
\def\caK{\mathcal{K}}
\def\scL{\mathscr{L}}
\def\frm{\mathfrak{m}}
\def\caO{\mathcal{O}}
\def\IP{\mathbb{P}}
\def\IR{\mathbb{R}}
\def\scS{\mathscr{S}}
\def\vd{\partial}
\def\frs{\mathfrak{s}}
\def\scX{\mathscr{X}}
\def\ZZ{\mathbb{Z}}
\let\SSS=\scriptstyle
\let\sss=\scriptscriptstyle
\let\ssm=\smallsetminus
\let\scSz=\scriptsize
\let\fnSz=\footnotesize
\newcommand{\FF}[2][n]{F^{\scriptscriptstyle(#1)}_{#2}}
\newcommand{\pFn}[2][]{{\Sigma^{#1}_{\!\smash{#2}}}}
\newcommand{\pDs}[1]{{\Delta^{\!\ast}_{\!\smash{#1}}}}
\newcommand{\pDN}[1]{{\Delta_{\!\smash{#1}}}}
\def\vC#1{\vcenter{\hbox{\hss#1\hss}}}
\def\TikZ#1{\begin{tikzpicture}#1\end{tikzpicture}}
\def\pM#1{\!\left(\mkern-2mu\begin{smallmatrix}#1\end{smallmatrix}\mkern-2mu\right)}
\def\bgK#1#2#3#4{{\def\arraystretch{.8}\arraycolsep=3pt%
                 \left#1\!\begin{array}{#2}#3\end{array}\!\right#4}}
\def\ssK#1#2#3#4{\text{\def\arraystretch{.3}\arraycolsep=2pt\scriptsize%
                 $\left#1\!\begin{array}{#2}#3\end{array}\!\right#4$}}
\makeatletter
\def\paragraph{\@startsection{paragraph}{4}{\z@}
           {.75ex \@plus.5ex \@minus.2ex}{-2mm}{\sf\bfseries\boldmath}}
\DeclareRobustCommand\widecheck[1]{{\mathpalette\@widecheck{#1}}}
\def\@widecheck#1#2{%
    \setbox\z@\hbox{\m@th$#1#2$}%
    \setbox\tw@\hbox{\m@th$#1%
       \widehat{%
          \vrule\@width\z@\@height\ht\z@
          \vrule\@height\z@\@width\wd\z@}$}%
    \dp\tw@-\ht\z@
    \@tempdima\ht\z@ \advance\@tempdima2\ht\tw@ \divide\@tempdima\thr@@
    \setbox\tw@\hbox{%
       \raise\@tempdima\hbox{\scalebox{1}[-1]{\lower\@tempdima\box\tw@}}}%
    {\ooalign{\box\tw@ \cr \box\z@}}}
\makeatother
\def\chX{{\mkern3mu\skew3\widecheck{\mkern-4mu\scX}}}
\def\chZ{{\skew3\widecheck{Z}}}
\def\tPX{{{\mkern1mu}^\triangledown\mkern-7mu\scX}}

\renewcommand{\leq}{\leqslant}
\renewcommand{\ge}{\geqslant}\renewcommand{\geq}{\geqslant}

   \parskip\medskipamount     \lineskip=0pt
\thicklines \footnotesep=2mm
\allowdisplaybreaks
\begin{document}

\setcounter{page}{1}
\thispagestyle{empty}

\begin{center}
\renewcommand{\thefootnote}{\fnsymbol{footnote}}
\setcounter{footnote}{0}

\vspace*{10mm}
{\Large\bfseries Beyond Algebraic Solutions to Stringy Spacetime%
\footnote{Based on the talk presented at the International Conference
{\em\/Nonlinearity, Nonlocality And Ultrametricity\/}
on the occasion of Branko Dragovich' 80th birthday,
26--30th May, 2025, Belgrade, Serbia}}\\*[5mm]
{\large\bf Tristan H\"ubsch}\\*
{\slshape Department of Physics that Astronomy, Howard University\\*[-2pt]
Washington, DC, 20059, USA;~ \texttt{thubsch@howard.edu}}\\*[10mm]
{\bf Abstract}\\[3mm]
\parbox{124mm}{Active feedback between geometry and physics is woven throughout the study of Nature at its fundamental level, and is of key importance in string theory. Methods of complex algebraic geometry in particular have brought about an unrivaled abundance of solutions, counted well into hundreds of orders of magnitude, reciprocated by the discovery of the wholly unexpected mirror symmetry. However, recent developments demonstrate that there are rich additional possibilities, made possible by certain generalizations that, at first glance, appear to be non-algebraic. Nevertheless, they are remarkably well-aligned within an overall mirror-symmetric framework, are amenable to almost as comprehensive quantitative analysis, and hint at a deeper relationship with symplectic geometry.}

\vspace*{10mm}
\begin{minipage}{124mm}\baselineskip=12pt plus1pt minus 1pt
\tableofcontents
\end{minipage}
\setcounter{footnote}{0}
\end{center}

\vfill

\section{Introduction, Rationale and Synopsis}
\label{s:IRS}
Already the very first concrete solution to Einstein's field equations~\cite{Schwarzschild:1916uq} made it clear that even ``empty'' (``sourceless,'' $T_{\mu\nu}\!=\!0$) observable spacetime can have a fascinating geometry, soon to be aptly reconstructed by algebraic methods~\cite{Kasner:1921Fin}, which also reveal its tantalizing 2-sheeted nature~\cite{rCF-BH}. Algebraic methods have continued to prevail in the study of (possible) spacetime geometry, in the last 42 years also within ``Calabi--Yau compactified'' string theory~\cite{rCHSW, rBeast2}, the past $\sfrac{1\!}3$-century development of which is firmly founded on the so-called gauged linear sigma model (GLSM)~\cite{rPhases, rMP0, Distler:1993mk, Schafer-Nameki:2016cfr}. Owing to their spatially extended nature, strings consistently propagate even through singular and variously defective spacetimes\cite{rSingS}, which then affords spacetime topology-changing phase transitions that may unify the admissible spacetime geometries into an irreducible ``worlds-wide web''\cite{rReidK0, rCYCI1, rGHC, rGHPT, rCGH1, rCGH2, rBeast2, Avram:1997rs, Avram:1995pu, Kreuzer:2000xy, rKS-CY, McNamara:2019rup}.

The underlying 1+1-dimensional worldsheet QFTs are routinely studied with at least $(0,2)$-super\-symmetry,\footnote{The worldsheet supersymmetry-complexified gauge symmetry is known to admit some mild degeneration or boundary sources (target-spacetime 'branes), relaxing further the underlying framework. In turn, the worldsheet quantum field theory itself has a manifestly $T$-dual and mirror-symmetric formulation, insuring a foundation to rely on\cite{Freidel:2015pka, Berglund:2021hbo}.} which complexifies the gauge group, $U(1)^n\hookrightarrow U(1;\IC)^n =(\IC^*)^n$, perfectly corresponding to the $(\IC^*)^n$-action of the (complex algebraic) toric geometry~\cite{rKKMS-TE1, rD-TV, rO-TV, rF-TV, rGE-CCAG, rCLS-TV, rCK}.
 This framework is home to the largest known database of some $\sfrac{1\!}2\!\times\!10^9$ {\em\/reflexive\/} polytopes, each encoding (weak/semi) Fano ($c_1\ge0$) {\em\/complex algebraic\/} toric 4-folds and their continuous families of anticanonical Calabi--Yau 3-fold hypersurfaces\cite{rKreSka00b}.
 Besides being connected by continuous (but singularizing) transitions and different (de)singularizations, this vast pool of constructions exhibits {\em\/mirror symmetry,} whereby the
$(\IC^*)^n$-equivariant (and quantum-corrected) cohomology rings of pairs of Calabi--Yau hypersurfaces, $(Z,\chZ)$, satisfy
$H^{p,q}(Z)\approx H^{n-p,q}(\chZ)$---{\em as rings,} equipped with the string-inspired quantum multiplication encoded by the Yukawa couplings in the full target-spacetime QFT physics~\cite{rGP1, rBH, rBaty01, rBatyBor1, rBH-LGO+EG, rCOK, rBatyBor2, rBatyBor3, Krawitz:2009aa, rMK-PhD, rLB-MirrBH, rACG-BHK, rF+K-BHK}.

In the remainder of this introduction, we review the structure of our millieu, the quantum field theory (QFT) framework of (super)string theory, and from the worldsheet vantage point: Here, we focus on worldsheet $(2,2)$-supersymmetric models with scalar superfields and abelian gauge symmetry, and especially the (abelian) gauged linear sigma models (GLSM)\cite{rPhases, rMP0, Distler:1993mk, Schafer-Nameki:2016cfr}.
Section~\ref{s:ShowX} presents the showcasing example, where the gauge $U(1;\IC)^2$-symmetry acting differently on an array of chiral superfields already results in an intricate field space, which in its ``geometric phase'' includes generalizations of Hirzebruch (rational, ruled) surface scrolls\cite{rH-Fm, rO-TV,rF-TV,rGrHa,rGE-CCAG,rCLS-TV,rKC-Fm}, and so generalizes them into the worldsheet QFT framework.
Section~\ref{s:AmbiDefo} presents the explicit deformation families of Hirzebruch surfaces of arbitrarily high twist, and
Section~\ref{s:DefoPic} then presents the explicit (2nd-tier) deformation families of Ricci-flat, i.e., Calabi--Yau hypersurfaces therein.
Our conclusions and outlook are collected in Section~\ref{s:CODA}, including some future aspirations in this direction.

\subsection{String Theory, from the Worldsheet Perspective}
\label{s:1.1}
In general, all physics modeling can be described in the framework outlined below\cite{rACPFT}, but here we focus on superstring models in particular.
\begin{description}[leftmargin=\parindent, topsep=-1pt]

\item[\bf Domain space, $\mathfrak{D}$]
 a genus-$g$ Riemann surface, $\Sigma_g$,
 ({\it i\/})~identified with the worldsheet swept out in time by the 
 propagating string, 
 ({\it ii\/})~equipped with a Lorentzian metric, $\gamma_{ij}(\xi)$,
 in terms of
 ({\it iii\/})~local coordinates, $\xi^i$.

\item[\bf Target space, $\mathfrak{T}$] a space, $\scX$,
 ({\it i\/})~identified with the spacetime through which the string 
 propagates,
 ({\it ii\/})~equipped with a Lorentzian metric, $G_{\mu\nu}(X)$,
 in terms of
 ({\it iii\/})~local coordinates $\langle X^\mu\rangle$,---{\em\/vacuum expectation values\/} (`vev's) of:
 
\item[\bf Immersion mapping] ``coordinate fields,''
 $X^\mu(\xi):\Sigma_g \to \scX$, which immerse the worldsheet in the target-spacetime and so depict which way and how the history of the propagating string unfolds.

\item[\bf Hamilton's action functional] 
 $S[X;\gamma;G,\dots]=\int_{\Sigma_g}\!\!\rd^2\xi~\scL$, with the Lagrangian density,
\begin{equation}
  \scL\big( X^\mu,(\vd_iX^\mu),\dots;
            \gamma_{ij}(\xi);G_{\mu\nu}(X), \dots \big),
 \label{e:CL}
\end{equation}
governs the dynamics/geometry of the string propagation through the spacetime $\scX$, depending on the above model-data, $\gamma,G,\dots$.
\end{description}
This structure determines the classical physics of string dynamics by minimizing Hamilton's action functional, i.e., by finding the ``classical'' immersions $X^\mu(\xi)$ that minimize $S[X;\dots]$, and so are ``on shell,'' i.e., satisfy the Euler-Lagrange equations corresponding to the action-minimizing requirement,
$\delta S[X;\dots]=0$.

The corresponding QFT {\em\/extends\/} this framework by noting:
\begin{enumerate}[itemsep=-1pt, topsep=-1pt]
 \item Only the initial and final configurations of the string, 
$X(\xi_i),X(\xi_f)\subset\scX$, are in fact observed and known, and specify the initial and final states.
 \item The worldsheet itself, $\Sigma_g$, connecting that initial to the final configuration is neither known nor observable---whence we must sum\footnote{This is the so-called {\em\/totalitarian principle\/}: “Everything not forbidden is compulsory.”\cite{rTHW-tot}.} over all of its possible $\Sigma_g$-shapes and 
 $(\xi;\gamma_{ij})$-parametrized sizes. These observations evidently relax the classical $\delta S[X;\cdots]=0$ requirement, consider the ensemble of {\em\/all\/} possible immersions, $X^\mu(\xi):\Sigma_g \to \scX$---each one {\em\/weighed\/} by the value of its classical action, $S[X;\cdots]$---and so define:
\end{enumerate}
\begin{description}[leftmargin=\parindent, topsep=-1pt]
 \item[\bf Path integral] or, sum-over-all-histories,
\begin{equation}
  Z[x;G,\kappa]:=\int\mkern-17mu\int\mkern-17mu\int\!\!\mathbf{D}[X]~
                  e^{-iS[X;\gamma;G,\dots]/\hbar}.
 \label{e:QZ}
\end{equation}
The formal integral over all immersions, $X^\mu(\xi):\Sigma_g \to \scX$, includes a sum over all genera, and the integration over the moduli (effective shape-and-size parameters) of arbitrary genus Riemann surfaces---which, surprisingly, {\em\/is\/} rigorously well defined via the so-called Deligne--Mumford compactification; see\cite{Harris:1987An-, Penner:1987The}.
\end{description}
The resulting ``quantization'' of the classical model~\eqref{e:CL} may be reformulated by the change of variables, $X^\mu\to X^\mu{-}x^\mu$, upon which the partition functional~\eqref{e:QZ} becomes a functional also of the chosen ``background value,'' $x^\mu:=\langle X^\mu\rangle$. In fortuitous circumstances (for so-called ``renormalizable'' models), we find that
\begin{equation}
  \int\mkern-17mu\int\mkern-17mu\int\!\!\mathbf{D}[X]~
                  e^{-iS[(X-x);\gamma;G,\dots]/\hbar}
  =e^{-iS_{\text{eff}}[x;\overline\gamma;\overline{G},\dots]/\hbar},
 \label{e:RZ}
\end{equation}
where $S_{\text{eff}}[x;\overline\gamma;\overline{G},\dots]$ is a Hamilton's action of the same form as the original action, 
$S[X;\gamma;G,\dots]$, merely with the model data, $\gamma,G,\dots$ ``deformed'' (renormalized) from their initial choices:
\begin{equation}
  \big\{ \gamma_{ij}(\xi); G_{\mu\nu}(X), \dots \big\} ~\leadsto~
  \big\{ \overline\gamma_{ij}(\xi); \overline{G}_{\mu\nu}(x),\dots \big\}.
 \label{e:RGE}
\end{equation}
Since the partition function~\eqref{e:QZ} is mostly computed perturbatively, the renormalization~\eqref{e:RGE} occurs in (iterative) steps, which is seen as a {\em\/flow,\/} towards a selection of data that no longer changes---which defines a ``renormalization fixed point'' in the model data space. 
\begin{rem} \label{r:2views}
As described above, this framework {\em\/presumes\/} a choice of a concrete target space(time), $\scX$, and equipped with a given metric,\footnote{\label{fn:other}For simplicity of exposition, the focus here is on the metric; this may well need to be and is straightforward to extend by other geometric/dynamic structures, such as complex structure, symplectic structure, gauge fields, torsion, etc.} so the only variable in the action functional is the mapping, i.e., $\scX$-vector of worldsheet fields. This is the ``target-spacetime vantage.''

By contrast, from the ``worldsheet vantage,'' one shifts gears by ``reverse-en\-gi\-neer\-ing'' the target spacetime: In this approach, the metric and other structures\footref{fn:other} and even topology\footnote{Topological (``nonlocal geometry'') information is typically encoded by way of boundary conditions and {\em\/quotienting,} i.e., by identifying different points of a space that are related by a symmetry group action. For example, a 2-torus is a finite planar parallelogram with periodic boundary conditions, but also the quotient $\IR^2/\Lambda$, where 
$\Lambda\subset\IR^2$ is a 2-dimensional lattice.} of the target spacetime, 
$\scX$, are {\em\/dynamical couplings\/} in the action (and partition) functional, which is in turn restricted in its form by presumed symmetries and overall self-consistency (anomaly-cancellation) conditions.\hrulefill
\end{rem}

\paragraph{Geometrostasis}
From this worldsheet vantage point, the (perturbatively computed) renormalization requirement for the model data---and the target space metric, $G_{\mu\nu}(X)$, in particular---to be at a renormalization fixed point, i.e., to be stable under ``quantum fluctuations'' was found by Daniel Friedan\cite{rF79a, rF79b} to exactly reproduces the Einstein equations,\footnote{This led to the introduction of the term ``geometrostasis'' in Ref.\cite{Braaten:1985is}, where Friedan's and some earlier work was generalized, in particular also to include geometric torsion.} in their Ricci form,
\begin{equation}
  0\overset!= \big[ R_{\mu\nu} - \tfrac{8\pi G_{\!N}}{c^4}
  \big(T_{\mu\nu}{-}\tfrac1{d-2}G_{\mu\nu}\,G^{\rho\sigma}T_{\rho\sigma}\big)
  \big] + O(\alpha'),
 \label{e:R=0}
\end{equation}
where $d=\dim(\scX)$, $\alpha'$ is the string worldsheet area characteristic size, and $T_{\mu\nu}$ is the energy-momentum density tensor for target-spacetime ``matter,'' i.e., all non-$G_{\mu\nu}(X)$ degrees of freedom. 
Notice that the ``trace-flipping'' of the energy-mo\-men\-tum density tensor itself depends on the spacetime metric, $G_{\mu\nu}$, and its inverse; the Ricci tensor of course depends on $G_{\mu\nu}(X)$, its derivatives and matrix-inverse.

The resulting conclusion, that
quantum stability of the worldsheet {\em\/quantum\/} field theory requires (to lowest $\alpha'$-perturbative order)
the target-spacetime observable metric, $G_{\mu\nu}$, to satisfy its {\em\/classical\/} equations of motion (Einstein's field equations) may be seen as a (worldsheet to target spacetime) layer-building generalization of Ehrenfest's theorem.
This notion in fact generalizes further, implying that quantum stability of more general worldsheet {\em\/quantum\/} field theory models induces---to lowest 
$\alpha'$-perturbative order---the standard, gauge interactions in target-spacetime, with subsequent quantum corrections inducing $O(\alpha')$-corrections to these a priori {\em\/classical\/} equations of motion, which one may regard as deformations. 

Having obtained classical target-spacetime equations of motion, it is then often straightforward to find a target-spacetime Hamilton's action\footnote{There do exist models with equations of motion that do not seem to be reproducible by any action-minimization principle, as well as models with a known partition functional for which no \eqref{e:RZ}-like description in terms of a classical action is known. I do not know how to extend the present discussion to include such models.} from whence these equations of motion follow via the usual variational calculus.
This endows string theory with a dynamics/geometry-governing ``layer-cake'' structure:
\begin{enumerate}[topsep=-1pt]

\item The worldsheet-layer (often superconformal\footnote{Conformal symmetry is required since worldsheet metrics differing by a local rescaling are physically indistinguishable. At least a worldsheet $(0,1)$-supersymmetry is included as the only {\em\/universal\/} method known to guarantee the removal of tachyonic states in string theory and so imply a stability of the ground states.}) quantum
 field theory model,
\begin{equation}
   \Big\{ (\mathfrak{D}\!=\!\Sigma_g,\gamma(\xi));~
          (\mathfrak{T}\!=\!\scX,G(X));~
          X\!:\Sigma_g\to\scX;~
          S[X^\mu;\, \gamma;\, G_{\mu\nu},\dots]\Big\},
 \label{e:wsh}
\end{equation}
 has a Schr{\"o}dinger-like picture, with a Hilbert space of states
 spanned by the (``2nd-quantization'') creation operators in $X^\mu$ (and their 
 superpartners), and parametrized by the ``couplings,'' $G_{\mu\nu},\dots$, as
 they turn up in the {\em\/computationally convenient\/} worldsheet action functional, $S(X;\gamma;G,\dots]$.

\item Quantum stability in this worldsheet-QFT model
 conditions (vanishing so-called $\beta$-functions) for the worldsheet couplings, $G_{\mu\nu},\dots$, may be reinterpreted as the
 classical equations of motion derivable from an action functional,
 $S[\phi;G_{\mu\nu};\mathfrak{g},\dots]$, with $\mathfrak{g}$ denoting
 the coupling parameters in this target-spacetime physics---all induced from the underlying worldsheet QFT, typically including also a characteristic energy-scale introduced via ``dimensional transmutation.''\footnote{Also, the Green's function in 1+1-dimensional spacetime of the worldsheet is logarithmic, thus perforce introducing a ``reference'' energy-scale.}

\item Understanding the classical fields ($G_{\mu\nu}(X),\dots$) to be
 vevs of (usually same-denoted) functions of quantum fields, let $\phi$ denote 
 quantum fluctuations about these classical ``background'' values, so the 
 dynamics of $\phi$ is governed by the target-spacetime coupling parameters, 
 $\mathfrak{g}$.

\item This induced target-spacetime model,
\begin{equation}
   \Big\{ (\mathfrak{D}\!=\!\scX,G(X));~
          (\mathfrak{T}\!=\!\mathscr{F},\mathfrak{g}(\phi));~
          \phi\!:\scX\to\mathscr{F};~
          S[\phi;G_{\mu\nu};\mathfrak{g},\dots] \Big\},
 \label{e:tgs}
\end{equation}
 may well then itself be ``quantized'' via its own \eqref{e:QZ}-like 
 path-integral. The quantum stability conditions for $\mathfrak{g}$ may then be 
 interpreted as classical equations of motion in the next layer, etc.
\end{enumerate}
Selectively fixing some of the couplings in the worldsheet model while allowing others to vary allows focusing on one or the other of the ``sectors'' in this layer-cake of QFTs, and provides for its versatility. In this vein, we will focus on modeling the compact (and presumed too small to be observed directly), complex 3-dimensional Calabi--Yau factor (denoted $\scX$), in the 9+1-dimensional spacetime, which is moreover independent of the (non-compact and big) observed spacetime, $\approx\IR^{1,3}$. That is, we assume the total, 9+1-dimensional stringy target-spacetime to be a rigid product, $\IR^{1,3}\!\times\!\scX\!$, and focus on studying $\scX\!$, to which we refer as (the non-trivial part of) the target-spacetime. Systematic modifications of this rigid Ansatz have been studied in the literature to various degree and extent; see Figure~1 and surrounding discussion in\cite{Berglund:2022qsb}, and the references therein.

\paragraph{Reverse-Engineering Target Spacetime.}
From this worldsheet point of view and with the logic of Remark~\ref{r:2views}, one considers QFT models built 
({\bf1})~on the domain of an a priori unspecified 1+1-dimensional, genus-$g$ Riemann surface, $\Sigma_g$. Summing over all versions of $(\Sigma_g,\gamma(\xi))$ makes all of them ``count.''
This affords including topology (worldsheet and target-spacetime) by way of specific boundary conditions, but here we focus on the local properties, so effectively work in 
$\Sigma_g\mapsto\IR^{1,1}$.

The dynamics of QFT models is then governed by
({\bf2})~the choice of a \underline{\em\/convenient\/} action functional, here constructed from
({\bf3})~``coordinate fields''\:=\:immersion mappings, $X^\mu(\xi)$, (these are chiral superfields, but we focus on the lowest component, complex scalar fields), which are subject to the action of a gauged $U(1;\IC)^n$ symmetry.
In this framework, the
({\bf4})~geometry (metric) of the target space---spanned by the ground state degrees of freedom---will only emerge through iterative computations, and as an interplay between the ``kinetic'' (also dubbed ``$D$-terms'') and ``interaction'' terms, also dubbed ``$F$-terms'' and ``mixed terms''; see~\eqref{e:U}, below.

In particular, it is the identifications under the gauge symmetry transformations that result in the $X^\mu(\xi)$-space being identified as a gauge $U(1;\IC)^2$-quotient---which is arguably a first indication of nontrivial topology. While ground states in this QFT are indeed determined as the minima of a certain potential (see~\eqref{e:U}, below), without the gauge-quotient, these minima would form a contractible cone of well-nigh trivial topology. The manyfold variations in gauge-quotienting, however, makes the myriads of possible potentials~\eqref{e:U} yield an embarrass of riches in the variety of topology and geometry of stringy spacetimes, which is at once the boon and the bane in string theory.

In turn, one of the marvels of the past decade or two is the increasingly accelerating development of various computer-intensive methods and techniques (neural networks, machine-learning, artificial intelligence,~\dots), which are now capable of computing and with increasing accuracy the metric (and other characteristic quantities\footnote{The rapidly growing literature on the subject is itself fascinating and would take us too far afield to provide an even remotely fair and functional review; suffice it here to direct the reader to the relatively recent works\cite{Berglund:2022gvm, Butbaia:2024tje, Constantin:2024yxh, Berglund:2024uqv, Constantin:2024yaz, Berglund:2024psp, Rahman:2026mfy} for starters.}) on Calabi--Yau manifolds hitherto analyzed without this explicit knowledge. These new computational methods are actively and radically reshaping the landscape of (not only) this endeavor in ways that are hard to map.
\dots{\em\/stay tuned.}

\subsection{Constrained QFT and GLSM}
\label{s:1.2}
To dynamically determine a target space, $\scX$, one must start {\em\/somewhere,} and---for conceptual but even more so computational convenience---one starts with a well-understood, well-known {\em\/ambient space,}\footnote{\label{fn:A}One seeks to work in a space where one knowns just about everything one needs for the purposes of computing the desired characteristics and properties of the subspaces of interest\cite{rBeast2}.} $A$, to serve as an auxiliary scaffolding framework within which one seeks to construct the Ricci-flat/Calabi--Yau space of our ultimate interest, $\scX$.
The ``{\em\/computationally convenient\/} action functional'' (in~\eqref{e:wsh} and step~({\bf1}) in the ``layer-cake'' outline above) then must reflect all the symmetries and properties of this chosen $A$, and also includes choices typically made for (computational) convenience.
This then limits---at the outset---the scope of such constructions, but also provides an avenue for subsequent extensions and generalizations, once one is aware of these structural details. Concretely in QFT with 1+1-dimensional domain spacetime, the standard lore has it that only scalar and spin-$\sfrac{1\!}2$ fields are dynamically interesting, as only they have propagating degrees of freedom. Merely to indicate further avenues of analysis, suffice it here to note that higher-spin fields in fact {\em\/do\/} contribute, primarily by imposing (gauge and otherwise) constraints, which in turn can be described using corresponding (lower-spin) {\em\/ghost\/} fields. Herein, we focus on the more basic level of discussion, as it will already hint at fascinating generalizations of the models most often used in the field and discussed in the past $\sfrac{1\!}3$-century literature.

\paragraph{The $\IP^n$ Template.}
For reference, recall that the best-known example of a convenient ambient space in which to construct spaces of interest as hypersurfaces is the complex projective space, $\IP^n$. Its QFT implementation~\cite{Eichenherr:1978SUN} starts with $n{+}1$ complex-valued fields, $X_i$, upon which a single $U(1)$ gauge symmetry acts, with equal charges $q(X_i)=q$. That is, among the $n{+}1$ separately defined $U_i(1)$-transformations, $X_i\to\lambda_iX_i$ with $\{0\neq\lambda_i\in\IC\}=:\IC^*=U(1;\IC)$, we {\em\/choose\/} to gauge the ``diagonal'' subgroup with $q(X_i)=q(X_j)=1$, leaving a remaining group of $U(1)^n$-transformations non-gauged, shown here in a conveniently chosen basis of $U(1)$-charges:
\begin{equation}
 \mkern60mu
 \begin{array}{@{}r@{~}r|cccccc@{}}
 &       &X_1 &X_2 &\cdots &X_n &X_{n+1} \\ \toprule
\multirow4*{\makebox[0pt][r]{\bf not gauged:}
            \makebox[0pt][l]{$\left\{\rule{0pt}{10mm}\right.$}}
 &   q_1 &1      &0      &\cdots &0      &-1\\ 
 &   q_2 &0      &1      &\cdots &0      &-1\\ 
 &\vdots &\vdots &\vdots &\ddots &\vdots &\vdots\\
 &   q_n &0      &0      &\cdots &1      &-1\\ \midrule 
\makebox[0pt][r]{\bf gauged:~}
 &q_{n+1}&1      &1      &\cdots &1      &1\\ 
 \end{array}
 \label{e:qPn}
\end{equation}
The first $n$ (row-wise) $n{+}1$-vectors are chosen orthogonal to the last one, \begin{equation}
  \sum_{i=1}^{n+1}q_a(X_i)\,q_{n+1}(X_i)=0,\quad a-1,\dots n,
 \label{e:MoriPn}
\end{equation}
so the non-gauged $U(1)^n$-transformations are orthogonal to the gauged $U_{n{+}1}(1)$. The gauge symmetry makes 
$\big(\lambda{\cdot}X:=(\lambda X_1,\cdots\lambda X_{n+1})\big)
 \approx(X_1,\cdots X_{n+1})$, forcing us to separate the $U(1)^{n+1}$-invariant subset, ${\bf E}_o:=\big((X_1,\cdots X_{n+1})=(0,\cdots0)\big)$, from its gauge-variant field-space complement,
${\bf E}^c_o:=\big\{(X_1,\cdots X_{n+1})\neq(0,\cdots0)\big\}$,
where the $U_{n+1}(1)$-transformation acts {\em\/freely\/}; there,
$\lambda{\cdot}X = {\lambda'}{\cdot}X$ holds precisely when
$\lambda = {\lambda'}$. 
This makes the gauge-quotient,
\begin{equation}
  \big\{\big((X_1,\cdots X_{n+1})\neq(0,\cdots0)\big) \approx
         (\lambda X_1,\cdots\lambda X_{n+1})\big\} =: \IP^n,
 \label{e:quoPn}
\end{equation}
well-defined. Also, $(0,\cdots0)$ is {\em\/unreachable\/} from any $X\neq0$ location by $\lambda\in\IC^*$-transformations, justifying the separation of the two $U_{n+1}(1)$-orbits.

A $\IP^n$-modeling QFT is constrained to a hypersurface,
$\{f(X)=0\}\subset\IP^n$, 
using a Lagrange multiplier, $\Lambda$, and a potential of the form
$\Lambda f(X)$. 
This constraining is well defined on the gauge-quotient space~\eqref{e:quoPn} 
precisely if $f(X)$ is homogeneous of degree $q_f=q_{n+1}\big(f(X)\big)$,
which is straightforward to generalize to supersymmetric QFT models of:
({\bf1})~complete intersections of hypersurfaces by including multiple constraints with multiple Lagrange multipliers, in
({\bf2})~products of projective spaces by including multiple separate 
$\IP^{n_k}$'s\cite{rChaSM, rMargD, rUDSS08, rUDSS09}; see also\cite{Witten:1987tv} and references therein. 
For~\eqref{e:qPn}--\eqref{e:MoriPn},
$U_{n+1}(1)$ gauge invariance of the constrained action functional forces $q(\Lambda)=-q_f$, so $\Lambda$ in fact must be a propagating field.\footnote{A propagator and a kinetic term would be radiatively introduced, were we to omit it.}
With the $U_{n+1}(1)$-charges~\eqref{e:qPn}, choosing $q\big(f(X)\big)=n{+}1$ implies
\begin{equation}
  \sum_iq_{n+1,i} + q_{n+1,\Lambda}=0,
 \label{e:CYL}
\end{equation}
and ensures that the gauged $U_{n+1}(1)$ (``fine structure'') coupling strength stays constant. This also implies
\begin{equation}\textstyle
  f(X) \in \big\{ \big(\bigoplus_{i=1}^{n+1} X_i\big)^{n+1} \big\} 
  \ni (\Pi X)\,{:=}\,\prod_iX_i,
 \label{e:K*Pn}
\end{equation}
where $(\Pi X)$ is the so-called ``fundamental monomial''\cite{rHY-SL2}, and the hypersurface, $\{f(X){=}0\}\subset\IP^n$ is Ricci-flat/Calabi--Yau\cite{rBeast2}. 
We close by noting that the non-gauged symmetries $U_1(1)\times\cdots U_n(1)$ with charges as in~\eqref{e:qPn} are all broken by a generic choice of $f(X)$ in~\eqref{e:K*Pn}, but that the so-called Fermat polynomial, $f_F(X)=\sum_i X_i^n$ preserves the discrete subgroup, with charges:
\begin{equation}
{\small
 \begin{array}{@{}r|cccccc@{}}
 U_a(1) &X_1 &X_2 &\cdots &X_n &X_{n+1} \\ \toprule\noalign{\vglue-2pt}
    q_1 &1      &0      &\cdots &0      &-1\\[-2pt]
    q_2 &0      &1      &\cdots &0      &-1\\[-5pt] 
 \vdots &\vdots &\vdots &\ddots &\vdots &\vdots\\[-1pt]
    q_n &0      &0      &\cdots &1      &-1\\[-2pt]\midrule\noalign{\vglue-1pt}
 q_{n+1}&1      &1      &\cdots &1      &1\\
 \end{array}
~~\to~~
 \begin{array}{@{}r|cccccc@{}}
 \ZZ_{n+1,a} &X_1 &X_2 &\cdots &X_n &X_{n+1} \\ \toprule\noalign{\vglue-2pt}
 \hat q_1 &1      &0      &\cdots &0      &n\\[-2pt]
 \hat q_2 &0      &1      &\cdots &0      &n\\[-5pt]
 \vdots &\vdots &\vdots &\ddots &\vdots &\vdots\\[-1pt]
 \hat q_n &0      &0      &\cdots &1      &n\\[-2pt]\midrule\noalign{\vglue-1pt}
 \hat q_{n+1}&1   &1      &\cdots &1      &1\\ 
 \end{array}
}
 \label{e:nZn+1}
\end{equation}
$\prod_{a=1}^n\big(\ZZ_{n+1,a}\subset U_i(1)\big)$, the ``diagonal subgroup'' of which is $\ZZ_{n+1}\subset U_{n+1}(1)$, generated by
$\hat{q}_{n+1,i}=\frac1{n+1}\sum_{a=1}^n[q_{a,i}~\mathrm{mod}\:(n{+}1)]$,\footnote{Note: $\sum_{a=1}^nq_a(X_{n+1})=n^2=(n^2{-}1){+}1=(n{-}1)(n{+}1){+}1
 \equiv1~~\mathrm{mod}\:(n{+}1)$.} and so is gauged. 
This leaves $(\ZZ_{n+1})^{n-1}$, generated by any $n{-}1$ linearly mod-$(n{+}1)$-independent combinations of $\hat{q}_1,\cdots\hat{q}_n$, as the maximal non-permutational symmetry group of Fermat hyper\-sur\-faces---which were key to discovering and constructing mirror models\cite{rGP1, rMirr01}.
\begin{rem}\label{r:Pn=TV}
The template~\eqref{e:qPn} already provides a first key link to complex-algebraic toric geometry~\cite{rKKMS-TE1, rD-TV, rO-TV, rF-TV, rGE-CCAG, rCLS-TV, rCK}: the non-gauged $U(1)$-charges of each field define a corresponding $n$-vector, 
$\skew{-1}\vec{q}(X_i)\mapsto\nu_i$, the $n{+}1$-tuple of which then generates the {\em\/fan\/} of $\IP^n$, realized as a toric variety.\hrulefill
\end{rem}

\paragraph{Worldsheet Supersymmetry and Mirror Symmetry:}
Integrating out (in principle) the gauge degrees of freedom and the Lagrange multiplier fields leaves a non-linear worldsheet QFT where the vevs of the remaining $X_i(\xi)$'s serve as a basis of local coordinates on the Ricci-flat/Calabi--Yau target space. The dynamics of these local coordinate fields, 
$X_(\xi)$, is governed by a {\em\/direct\/} ({\em\/intrinsic\/}) action functional featuring the standard kinetic term,
$\gamma^{\alpha\beta}(\xi)\,(\vd_\alpha X^\mu)(\vd_\beta X^\nu)\,G_{\mu\nu}(X)$,
where $G_{\mu\nu}(X)$ is the target space Ricci-flat/Calabi--Yau metric. This is guaranteed to exist by Yau's celebrated theorem\cite{rYau77a, rYau77b}, but is not known in closed form in essentially any compact example relevant to string theory---which is the key reason for the past four decades of studying string theory models in {\em\/indirect\/} ways, via constrained models, and in particular the GLSM to be discussed hereafter.

The {\em\/direct\/} description is however indispensable in making manifest the naturalness of a key, hallmark property, especially so with assumed worldsheet $(2,2)$-supersymmetry: This supersymmetry 
induces a target-space supersymmetry rendering the immersion mapping ``coordinate fields'' complex, $X_i(\xi)\to X^\mu(\xi),X^{\bar\mu}(\xi)$, and
pairs them with two superpartners,
$\psi^\mu_{\pm}(\xi),\psi^{\bar\mu}_{\pm}(\xi)$. 
A dynamically independent half\footnote{A maximal collection of fermionic fields the mutual Poisson brackets of which all vanish.} of these fermionic superpartners provide a local basis for the local tangent space; the remaining (canonically conjugate) half spanning the canonically dual cotangent space. Whereas $\gamma^{\alpha\beta}(\xi)$ is the worldsheet (inverse) metric, over which the QFT partition functional is integrated, the components of $G_{\mu\nu}(X)$ serve as coupling parameters in such worldsheet QFT models, for which~\eqref{e:R=0} specifies the renormalization fixed-points and geometrostasis.

On a complex (factor of the) target spacetime (which exhibits worldsheet $(2,2)$-supersymmetry), $\scX$, this fermion-halving may be chosen, e.g.\ (see~\cite{rHSS} and references therein),
\begin{subequations}
 \label{e:SuSyGeom}
\begin{alignat}9
&&G_{\mu\bar\nu}(X)\,\psi_+^{\bar\nu}(\xi) &&=:\psi_{+\mu}(\xi)
  &\mapsto\vd_\mu \quad&\text{and}\quad \psi_+^\mu(\xi)&\mapsto\rd X^\mu,\\
  \text{vs.}&\quad
 &G_{\mu\bar\nu}(X)\,\psi_-^\mu(\xi) &&=:\psi_{-\bar\nu}(\xi)
  &\mapsto\vd_{\bar\nu} \quad&\text{and}\quad \psi_-^{\bar\nu}(\xi)&\mapsto\rd X^{\bar\nu}.
\end{alignat}
\end{subequations}
This maps the Fock-space elements to Dolbeault cohomology groups:
\begin{subequations}
 \label{e:SuSyCoho}
\begin{alignat}9
  h_{\bar\nu_1\cdots\bar\nu_q}^{\mu_1\cdots\mu_p}(X,\bar X)\,
  \psi_-^{\bar\nu_1}{\cdots}\psi_-^{\bar\nu_q}\,
   \psi_{+\mu_1}{\cdots}\psi_{+\mu_p}
  |0\rangle &\mapsto H^q_{\sss\bar\vd}(\scX\!,\wedge^pT)
 &&=H^{n-p,q}_{\sss\bar\vd}(\scX\!,\mathcal{K}^*), \label{e:n-p,q}\\
  \omega_{\mu_1\cdots\mu_p\,\bar\nu_1\cdots\bar\nu_q}(X,\bar X)\,
  \psi_-^{\bar\nu_1}{\cdots}\psi_-^{\bar\nu_q}\,
   \psi_+^{\mu_1}{\cdots}\psi_+^{\mu_p}
  |0\rangle &\mapsto H^q_{\sss\bar\vd}(\scX\!,\wedge^pT^*)
  &&=H^{p,q}_{\sss\bar\vd}(\scX), \label{e:p,q}
\end{alignat}
\end{subequations}
where $\mathcal{K}^*_{\scX}:=\wedge^nT_\scX$ is the anticanonical bundle of the tanrget space, $\scX$, and the rightmost equality in~\eqref{e:n-p,q} exhibits the identity
\begin{equation}
   \wedge^pT\overset{\sss\rm id}{=}\wedge^{n+(p-n)}T=
  \mathcal{K}^*_{}{\otimes}\wedge^{n-p}T^*.
\end{equation}
The assignments~\eqref{e:SuSyGeom} and~\eqref{e:SuSyCoho} make it clear that complex-conjugating only the 
$\psi_+\leftrightarrow\bar\psi_+$ swaps $\wedge^pT\leftrightarrow\wedge^pT^*$, which exhibits the elementary origin of {\em\/mirror duality\/} in worldsheet $(2,2)$-supersymmetric models\cite{Polchinski:1998rr, rHSS}: 
$\!\chX$ is the mirror of $\scX$ if the cohomology rings\footnote{This statement of mirror duality presumes the ``quantum'' (deformation of the wedge) ring structure of these cohomology groups, the structure constants defining target-space Yukawa couplings~\cite{Kontsevich:1995wkA, Kontsevich:1994Mir, rCK, rD+G-DBrM}; their dimensions of course must agree regardless of the ring structure.} satisfy
$H^q_{\sss\bar\vd}(\scX\!,\wedge^pT)\approx 
 H^q_{\sss\bar\vd}(\chX,\wedge^pT^*)$ and vice versa, for all $p=0,\dots\dim(\scX)$. 
Comparison of the ultimate, rightmost equalities in~\eqref{e:n-p,q} and in~\eqref{e:p,q} exhibits, in turn, that in target-spaces with a trivial (anti)canonical class, $\mathcal{K}^*_{\scX}\,{=}\,\mathcal{O}_\scX$ so 
$c_1(\scX)\,{=}\,0$ and $\scX$ is Ricci-flat/Calabi--Yau,
{\em\/mirror duality\/} acts a reflection in the Hodge-decomposed cohomology 
{\em\/ring,\/} $H^{n-p,q}(\scX)\approx H^{p,q}(\chX)$.

\paragraph{GLSM}
In some ways a straightforward adaptation of constrained (nonlinear) gauged sigma models, the {\em\/gauged linear sigma model\/} (GLSM)~\cite{rPhases, rMP0, Distler:1993mk, Schafer-Nameki:2016cfr} provides a very large class of worldsheet theories, originally formulated with worldsheet $(2,2)$-super\-sym\-metry to facilitate the analysis, and which we will assume throughout this discussion. In some of the simplest cases, their low-energy limits straightforwardly include the nonlinear, constrained projective-space models discussed above, and so generalize them.

In their (simplest) worldsheet $(2,2)$-supersymmetric incarnation, GLSMs are well specified by providing the following:
\begin{enumerate}[topsep=-1pt]

\item A list of $n{+}r$ chiral $(2,2)$-superfields, $X_i$, the vevs of the lowest 
 (bosonic, complex scalar) component fields of which provide {\em\/homogeneous\/} 
 coordinates for the (complex $n$-dimensional) compact factor in the target 
 spacetime.

\item A list of $r$ twisted-chiral $(2,2)$-superfields, $\sigma_a$, in 1--1
 correspondence with the gauged $U(1)^r$ symmetry. (Gauge field 2-vector
 potentials on the 1+1-dimensional worldsheet have no physical degrees of 
 freedom, but each is accompanied by a complex scalar field, also denoted 
 $\sigma_a$.)

\item The worldsheet-supersymmetric action of the $U(1)^r$ gauge symmetry
 automatically becomes complexified, $U(1)^r\to U(1;\IC)^r=(\IC^*)^r$, which
 acts by nonzero complex rescaling,
 $X_i\to\lambda{\cdot}X_i:=\prod_a\lambda_a^{q_{ai}}X_i$, where
 $q_{ai}=q_a(X_i)$ is the charge of $X_i$ with respect to the 
 $a^\text{th}$ $U(1)$, with $\lambda_a$ nonzero complex number-valued chiral 
 superfields: $\lambda_a\in\IC^*$, where $\IC^*:=(\IC\smallsetminus\{0\})$.
\end{enumerate}
The Lagrangian super-density consists of the standard, gauge $U(1)^r$-invariant kinetic term and a superpotential of the general form 
$\sum_{\alpha=1}^K \Lambda_\alpha\,f_\alpha(X)$, where we focus on the simplest, $K=1$ case, and write $\Lambda_1=X_0$. The function $f(X)$ is chosen quasi-homogeneous, $f(\lambda{\cdot}X)=\prod_a\lambda^{q_{af}}f(X)$, and $q_a(X_0)=q_{a0}$, so that the superpotential $X_0{\cdot}f(X)$ would be $U(1)^r$-invariant:
\begin{equation}
  \sum_iq_{ai} + q_{a0}=0,\quad a=1,\cdots r.
 \label{e:CY}
\end{equation}
This condition simultaneously:
({\bf1})~insures general gauge invariance for the superpotential~\eqref{e:CY},
({\bf2})~guarantees the anomaly cancellation for each gauge $U(1)$ symmetry,
and is also 
({\bf3})~the condition for the hypersurface
$\{f(x)=0\}\subset\big(\{X_i\}/U(1)^r\big)$ to have a vanishing 1st Chern class, and so admit a Ricci-flat K{\"a}hler metric---i.e., to be a Calabi--Yau hypersurface.
({\bf4})~The condition~\eqref{e:MoriPn} in fact also implies an array of ``biangle''\footnote{In 1+1 dimensional QFT, the analogue of the Adler-Bell-Jackiw 1+3-dimensional QFT anomaly computed by Feynman ``triangle diagrams'' involves {\em\/biangles\/}---internal loops coupling to two external currents, whence the bi-linear condition~\eqref{e:MoriPn}.} mixing-anomaly cancellations.
This triple/quadruple implication of the condition~\eqref{e:CY} firmly correlates QFT and geometric aspects of the analysis.

In the low-energy regime, the $X_0$ field indeed behaves as a Lagrange multiplier,\footnote{To be precise, that role is played by the 
$\int\!\rd^2\theta\,X_0$ auxiliary component in the chiral superfield. The equation of motion for this field is {\em\/always\/} algebraic, and so guarantees the enforcing of the $f(X)=0$ constraint regardless of the dynamics of the (rest of) the $X_0$ superfield.} enforcing the constraint $f(X)=0$---in the so-called ``geometric phases'' (see below)---which also restricts the $A$-``coordinate fields,'' $X_i$, and their superpartners to local $\scX$-coordinates and (co)tangent vectors;
integrating out the gauge degrees of freedom replaces the standard bilinear gauge-invariant kinetic term with the non-linear sigma model with the $\scX$-local metric\cite{rPhases, rBeast2}; see\cite{McOrist:2011bn, Bertolini:2017lcz, Melnikov:2019tpl, Ashmore:2019rkx, Sharpe:2024dcd} for more recent and more general (only $(0,2)$-supersymmetric) discussion. 
This provides a general overview of how these models generate the geometry of the so-described target spaces, to be regarded as hypersurfaces $\{f(X)=0\}\subset A$---which provides the technical ease bias towards ``well known'' ambient spaces, $A$; see footnote~\ref{fn:A}.

After integrating out the ``auxiliary fields'' (the equations of motion of which are algebraic, non-dynamical), the effective worldsheet $(2,2)$-supersymmetric Lagrangian density contains the potential that is the sum of positive semi-definite terms:
\begin{equation}
 \underbrace{\sum_a\!\Big(\!\sum_iq_{ai}|X_i|^2{-}t_a\!\Big)^2}_{D\text{-terms}}
 ~+~\underbrace{|f(X)|^2
  {+}|X_0|^2\sum_i\big|\frac{\vd f}{\vd X_i}\big|^2}_{F\text{-terms}}
 ~+~\underbrace{\sum_{a,b}\overline\sigma_a\sigma_b\sum_iq_{ai}q_{bi}|X_i|^2}_{\text{mixed terms}}.
 \label{e:U}
\end{equation}
The ground state is therefore determined by the vanishing of each summand, which provides a coupled system of algebraic equations that are usually analyzed in turn, as provided in~\eqref{e:U}. We will do so in detail, in a concrete, showcasing example and following\cite{rBH-gB}, rather than attempting to completely specify a general, and invariably notationally forbidding algorithm.
\begin{rem}\label{r:noW}
Furthermore, by completely omitting the superpotential---which is a choice preserved through renormalization in worldsheet $(2,2)$-supersymmetric QFT---this GLSM simply describes the embedding space as the gauge-quotient, $\{X_i\}/U(1)^r$. Thus, a GLSM may be thought of as a {\em\/workshop\/} in which to study both ``ambient spaces'' with hallmark $(U(1)=S^1)^{n+r}$-transformations (``actions,'' in mathematical literature) as well as hypersurfaces in them, identified as ground states determined by the minima of the effective (now included) potential.\hrulefill
\end{rem}
We next turn to address the ambient spaces first---and by switching to the concrete showcasing sequence of examples for notational simplicity.

\section{A Showcasing Example Sequence}
\label{s:ShowX}
Following primarily Ref.\cite{rBH-gB}, and with the subsequent works\cite{Berglund:2022dgb, Berglund:2024zuz, Hubsch:2025sph, Hubsch:2025teh, Hubsch:2026wxe} in mind, consider the rather straightforward generalization of~\eqref{e:qPn} and Remark~\ref{r:Pn=TV}:
\begin{equation}
 {\small\hspace{10mm}
  \begin{array}{r|@{~}r|rrr@{~}|@{~}rr@{~}|}
 \boldsymbol{n=3} & {\bf0} &\vec\nu_1 & \vec\nu_2
  & \vec\nu_3 & \vec\nu_4 & \vec\nu_5\\ \toprule
  \multirow3*{\raisebox{13pt}{\makebox[0pt][r]
              {$\pFn{\FF[3]m}\left\{\rule{0mm}{6mm}\right.$}}}
      &0 &-1 & 1 & 0 & 0 & -m \\[-2pt]
      &0 &-1 & 0 & 1 & 0 & -m \\[-2pt]
      &0 & 0 & 0 & 0 & 1 & -1 \\ \hline
 \text{(added)}
      &1 & 0 & 0 & 0 & 0 &  0 \\[-2pt] \midrule \noalign{\vglue-2pt}
  \multirow4*{\raisebox{18pt}{\makebox[0pt][r]
              {gauged $U(1)^2\left\{\rule{0mm}{8mm}\right.$}}}
  q_1 & -3 & 1 & 1 & 1 & 0 & 0 \\[-2pt]
  q_2 & (m{-}2) &-m & 0 & 0 & 1 & 1 \\[-2pt]
  q_3 & -2(m{+}1) &0 & m & m & 1 & 1 \\[-2pt]
  q_4 & 0 &-2(m{+}1) &(m{-}2) &(m{-}2) & 3 & 3 \\ \bottomrule
     & X_0 & X_1 & X_2 & X_3 & X_4 & X_5\\
  \end{array}}
 \label{e:q1-4}
\end{equation}
where we focus on the $n=3$ case of so-called Hirzebruch scrolls, $\FF[3]m$.
As above, the first four rows of $U(1)$-charges provide a basis for the non-gauged symmetry group, while any two of the remaining four rows provide a charge-basis for the gauged $U(1)^2$ symmetry.
This encodes the toric $U(1;\IC)^2\approx(\IC^*)^2$-action.
The $q_1,\cdots q_4$-row entries satisfy the key (Calabi--Yau) requirement~\eqref{e:CY}, $\sum_{i=0}^{n+2}q_a(X_i)=0$, and are chosen so as to vanish for (at least) one of the $x_i$. As 6-vectors, any two of the four $q_a$ are linearly independent and provide a priori equally valid choices (basis) for the $U(1)^2$-charge assignments in the GLSM.

In turn, the (complex) chiral GLSM superfields, $X_i$, and their lowest component fields correspond to the $n$-component column-vectors assembled from the upper rows in~\eqref{e:q1-4}. They generate the toric fan, $\pFn{\FF[3]m}$, and completely encode this ambient toric variety, called a Hirzebruch scroll, $\FF[3]m$, generalizing the well-known Hirzebruch surfaces~\cite{rH-Fm, rGrHa}; see also Refs.~\cite{rO-TV,rF-TV,rGrHa,rGE-CCAG,rCLS-TV,rKC-Fm}.
Generalizing~\eqref{e:MoriPn}, these fan-generating $n$-vectors correlate with the gauged $U(1)$ symmetry charges (lower rows in~\eqref{e:q1-4}) by satisfying the condition:
\begin{equation}
  \sum_{i=0}^{n+2}q_a(X_i)\,\vec\nu_i=0,\quad \text{for all}~~a,
 \label{e:q.nu}
\end{equation}
as well as that
the $\vec\nu_1,\cdots \vec\nu_5$ column 3-vectors of the
the first three rows and any two of the $q_a$-rows in~\eqref{e:q1-4}
form a $5\!\times\!5$ regular matrix (with determinant $-(6{+}2m^2)$), i.e., these five rows and columns form five linearly independent 5-vectors.
Furthermore, the first $3{+}1$ rows of $\nu_0,\cdots\nu_5$ together with any two $q_a$-rows in~\eqref{e:q1-4} stacked underneath form a regular $6{\times}6$ matrix, so these six rows and six columns form six linearly independent 6-vectors.\footnote{For $\FF{m}$, the determinant of both of these matrices is 
$(-1)^{n+1}\big(2n{+}(n{-}1)m^2\big)$.}

Notably, standard (complex-algebraic) toric geometry only uses $q_1,q_2$, the so-called {\em\/Mori vectors,} since their non-negative linear combinations reproduces $q_3=mq_1+q_2$.
Albeit standard in complex-algebraic toric geometry, this non-negativity does not seem to have any relevance to the worldsheet GLSM.
Also, the $q_4$ choice cannot turn up in standard (complex-algebraic) toric geometry\cite{rO-TV, rCLS-TV, rBKK-tvMirr}, but will turn up in the GLSM analysis below.

The 2-dimensional version of this same toric variety was in fact originally defined in 1951\cite{rH-Fm} as the {\em\/bi-projective hypersurface,}
\begin{equation}
  \FF{m}:=\big\{p_0(x,y):= x_0\,y_0^m+x_1\,y_1^m=0\big\}\subset\IP^n\times\IP^1,
 \label{e:nFm}
\end{equation}
which extends straighforwadly to all $n\geq2$,
and we will use this correspondence, notably also because constraining (products of) $\IP^n$ models is a well studied QFT framework~\cite{Eichenherr:1978SUN, rChaSM, rMargD, rUDSS08, rUDSS09}. In addition, Hirzebruch scrolls are also known as
the {\em\/projective bundles,}
  $\FF{m}:=\IP\big(\caO_{\IP^1}(m)\oplus\caO_{\IP^1}^{\oplus(n-1)}\big)$ over the base-$\IP^1$,
and as
the {\em\/$m$-twisted $\IP^{n-1}$-bundles over $\IP^1$.}
The widespread and manyfold study of these varieties within diverse subfields of algebraic geometry motivates their use. Their multiple equivalent definitions provide alternative ways to calculate desirable information, to which end a coordinate-level identity between the bi-projective embedding~\eqref{e:nFm} and the toric specification~\eqref{e:q1-4} will be useful; see \S\:\ref{s:2P-TV} and Refs.\cite{Berglund:2022dgb, Berglund:2024zuz, Hubsch:2025sph, Hubsch:2025teh, Hubsch:2026wxe}.

\subsection{Gauge Quotient Partitioning}
Since the field space spanned by the $X_i$ is in fact a $U(1;\IC)^2$-gauge quotient, and the $U(1;\IC)^2$-action~\eqref{e:q1-4} on 
$(X_0;X_1,\cdots X_5)\in\IC^6$ is evidently not uniform, one must apportion ({\em\/partition\/}) the affine field-space, $\IC^6$, into regions of uniform $U(1;\IC)^2$-action to have a well-defined quotient. 

The $U_a(1)$ gauge transformations~\eqref{e:q1-4} clearly partition the six variables, $X_i$, into four groups,
\begin{equation}
{\small
\begin{array}{@{}r@{\:}|@{\:}c@{\:}|@{\:}c@{\:}|@{\:}c@{\:}|@{\:}c@{}}
 \textbf{Factor} & \IC^1_0\!=\!\{X_0\} &\IC^1_1\!=\!\{X_1\}
  &\IC^{n-1}_2\!=\!\{X_2,\cdots X_n\} &\IC^2_3\!=\!\{X_{n+1},X_{n+2}\}\\*[2pt]\toprule
 \textbf{Fixed by} & U_4(1;\IC) & U_3(1;\IC) &U_2(1;\IC) & U_1(1;\IC)\\ \bottomrule
 \textbf{Fixed set} & {\bf E}_0=(0) & {\bf E}_1=(0)
  &{\bf E}_2=(0,\cdots 0) & {\bf E}_3=(0,0)\\
\end{array}}
 \label{e:theEs}
\end{equation}
Subsets of the field space fixed by more than one of the $U_a(1)$ gauge transformations~\eqref{e:q1-4} are easily defined, such as
\begin{equation}
  {\bf E}_{23}=\{X_2,\cdots X_n=0 ~\&~ X_{n+1},X_{n+2}=0\},\quad\text{etc.}
\end{equation}
Complements are defined as
\begin{equation}
  {\bf E}_0^c=\{X_0\neq0\}=(\IC^1_0\smallsetminus{\bf E}_0),\quad
  {\bf E}_1^c=\{X_1\neq0\}=(\IC^1_1\smallsetminus{\bf E}_1),\quad
  \text{etc.}
\end{equation}

With this partitioning, the $U(1;\IC)^2$ gauge symmetry defines:
\paragraph{I.}
Omitting $X_0$ and considering the $U_1(1;\IC)$- and $U_2(1;\IC)$-transformations:
\begin{equation}
 \Big(\underbrace{\big((\IC^n_{12}\ssm{\bf E}_{12})/U_1(1;\IC)\big)}
      _{=\,\IP^{n-1}_\text{fiber}}~\times~
      \big(\IC^2_3\ssm{\bf E}_3\big)\Big)\,\Big/\,U_2(1;\IC)
  ~\simeq~ \FF{m}
 \label{e:nFmQuot}
\end{equation}
where $U_2(1;\IC)$ projectivizes $\IC^2_3\ssm{\bf E}_3\to\IP^1_\text{base}$ while also $m$-twisting\footnote{\label{n:mTw}The $U_2(1)$-transformation is trivial on 
$\IP^1_{\sss\text{base}}\ni(X_{n+1},X_{n+2})\simeq
 (\lambda\,X_{n+1},\lambda\,X_{n+2})$, but is a non-trivial coordinate reparametrization on 
$\IP^{n-1}_{\sss\text{fiber}}\ni(X_1,X_2,\cdots X_n)\not\simeq
 (\lambda_2^{-m}\,X_1,X_2,\cdots X_n)$.} $\IP^{n-1}_\text{fiber}$. This defines an $m$-twisted $\IP^{n-1}_\text{fiber}$-bundle over $\IP^1_\text{base}$, the Hirzebruch $n$-fold scroll, $\FF{m}$.

\paragraph{II.}
Considering instead the $U_2(1\;\IC)$- and $U_3(1;\IC)$-transformations provides:
\begin{equation}
 \Big((\IC^1_1\ssm{\bf E}_1)~\times~
      \underbrace{\big((\IC^{n+1}_{23}\ssm{\bf E}_{23})/U_3(1;\IC)\big)}
      _{=\,\IP^n_{(m:\cdots:m:1:1)}}\Big)\,\Big/\,U_2(1;\IC),
 \label{e:buWCP}
\end{equation}
where $U_2(1;\IC)$ projectivizes $\IC^1_1\ssm{\bf E}_1\to\IP^1_{\text{exc.}}$ along $(X_2,\cdots X_n;\,0,0)$, which is $U_2(1)$-invariant and is fixed by 
$\ZZ_m\subset U_3(1)$. Indeed, $(X_2,\cdots X_n;\,0,0)$ is the $\ZZ_m$-singular locus in the weighted-projective space, $\IP^n_{\!\sss(m:\cdots:m:1:1)}$, here desingularized by replacing each point by a copy of the $X_1$-parametrized ``exceptional set,'' 
$\IP^1_{\text{exc.}}$.
\begin{rem}\label{r:U1-U3}
Since the $q_2,q_3$-basis~\eqref{e:q1-4} is equivalent to the $q_1,q_2$-basis, the iterated quotient~\eqref{e:buWCP} is isomorphic to~\eqref{e:nFmQuot}, thus giving the Hirzebruch $n$-fold, $\FF{m}$, another alternative description, as the MPCP-desingularization of $\IP^n_{\!\sss(m:{\cdots}:m:1:1)}$~\cite{rBaty01}.\hrulefill
\end{rem}
Replacing a complement $(\IC^d_{\cdots}\ssm{\bf E}_{\cdots})$ of a fixed set with the fixed set ${\bf E}_{\cdots}$ itself, and combinatorially in the various factors of~\eqref{e:nFmQuot} and~\eqref{e:buWCP}, yields the ``complementary'' partition in the field space:
\paragraph{III.}
The Landau-Ginzburg partition:
\begin{equation}
  [F_m^{\sss(n)}]\strut_{\text{LGO}}:~
  \Big((\IC^n_1\ssm{\bf E}_1)~\times~
         \big(({\bf E}_{23})/U_3(1;\IC)\big)\Big)\,\Big/\,U_2(1;\IC),
\end{equation}
\paragraph{IV.}
The hybrid partition:
\begin{equation}
  [F_m^{\sss(n)}]\strut_{\text{hyb.}}:~
  \Big(\big[({\bf E}_{12})/U_1(1;\IC)\big] ~\times~
      \big(\IC^2_3\ssm{\bf E}_3\big)\Big)\,\Big/\,U_2(1;\IC).
\end{equation}

The (complexified) $U(1;\IC)^2$ gauge symmetries apportion the field-space to the $U(1;\IC)^2$-equivalence classes in the following four field-space regions:
\begin{equation}
 \begin{array}{c@{\qquad\qquad}c}
   \makebox[0pt][r]{phase~IV~=~~}
   \big[{\bf E}^c_0\times{\bf E}_{12}\times{\bf E}^c_3\big]
  &\big[{\bf E}_0\times{\bf E}^c_{12}\times{\bf E}^c_3\big]\makebox[0pt][l]{~~=~phase~I}\\*[8mm]
  \TikZ{\path[use as bounding box](0,0);
    \draw[blue](.35,.35)tonode[left=-2pt]{$\SSS0$}++(0,.8);
    \draw[blue](2.05,.35)--node[below left=-2pt]{$\SSS2$}++(-.6,.8);
    \draw[blue](4.65,.35)--node[left=-2pt]{$\SSS0$}++(0,.8);
    \draw[blue](5.6,.35)--node[left=-2pt]{$\SSS1$}++(0,.8);
    \draw[blue](6.5,.35)--node[above right=-2pt]{$\SSS2$}++(-.8,.8);
    \draw[blue](6.6,.35)--node[right=-2pt]{$\SSS3$}++(0,.8);
    \draw[thick,red,densely dotted,-stealth](1.25,.35)
       --node[left=-2pt]{$\SSS1$}++(0,.8);
    \draw[thick,red,densely dotted,stealth-](2.2,.35)
       --node[right=-2pt]{$\SSS3$}++(.1,.8);
    \draw[thick,red,densely dotted,-stealth]
       (.35,1.7)to++(0,.2)to++(4.25,0)to++(0,-.25);
    \draw[blue](2.3,1.6)to++(0,.4)to++(4.3,0)to++(0,-.4);
    \draw[thick,red,densely dotted, stealth-]
       (1.25,1.65)to++(0,.45)to++(4.35,0)to++(0,-.45);
    \draw[thick,red,densely dotted,-stealth]
       (.35,-.1)to++(0,-.2)to++(4.3,0)to++(0,.2);
    \draw[blue](1.25,-.1)to++(0,-.3)to++(4.35,0)to++(0,.3);
    \draw[thick,red,densely dotted, stealth-]
       (2.15,-.1)to++(0,-.4)to++(4.35,0)to++(0,.35);
            }
  \makebox[0pt][r]{phase~III~=~~}
  \big[{\bf E}^c_0\times{\bf E}^c_1\times{\bf E}_{23}\big]
  &\big[{\bf E}_0\times{\bf E}^c_1\times{\bf E}^c_{23}\big]\makebox[0pt][l]{~~=~phase~II} \\[1mm]
 \end{array}\rule[-12mm]{0mm}{27mm}
 \label{e:EvsP}
\end{equation}
where the $\IC^1_0$ factor is now also included.
 The (red) dotted arrows indicate the (dimensional) collapse ${\bf E}^c_i\to{\bf E}_i$, from the complement of the (gauge-fixed) exceptional set to the exceptional set itself. The plain (blue) solid lines trace unchanged factors in the tensor product.
The four distinct regions are labeled as ``phases'' (adopting the by now standard nomenclature\cite{rPhases}).

As in Ref.~\cite{rPhases}, the variable $X_0$ is seen to provide a fiber-coordinate for the degree-$\pM{n\\2{-}m}$ line bundle over $\FF{m}$, which is identified as $\caK^*_{\smash{\FF{m}}}$, the anticanonical bundle of $\FF{m}$. The zero-locus of each of its sections is a Calabi--Yau hypersurface.

\subsection{VEVs and Phases}
Having apportioned the field space into $U(1)^2$-domains, consider now the possible choices of ground-state vev's, which maintain the vanishing of the potential~\eqref{e:U}. In particular, with the $(q_1,q_2)$-basis in~\eqref{e:q1-4}, the vanishing of the ``D-terms'' implies:
\begin{subequations}
 \label{e:VEVD}
\begin{alignat}9
0=& \Big({-}n|X_0|^2{+} \sum_{i=1}^n|X_i|^2 -t_1\Big), 
 \label{e:VEVD1}\\
0=& \Big((m{-}2)|X_0|^2 {-}m |X_1|^2+\sum_{j=1}^2|X_{n+j}|^2 -t_2\Big),
 \label{e:VEVD2}
\end{alignat}
\end{subequations}
The ``Fayet-Iliopoulos'' coupling-plane, $(t_1,t_2)$, effectively parametrizes the $U(1)^2$ symmetry breaking patterns, and the following demarcations emerge\cite{rMP0, rBH-gB}---fully consistent with~\eqref{e:EvsP} above:
\paragraph{(i)}
 Along ${\bf E}_{012}$, where $X_0,{\cdots},X_n=0$ but 
 $(X_{n+1},X_{n+2})\neq(0,0)$, $U(1)^2\to U_1(1)$;
 $\mathrm{lcm}[q_2(X_{n+1},X_{n+2})]=1$ breaks $U_2(1)$ completely. 
 In this case,~\eqref{e:VEVD1} sets $r_1=0$ and~\eqref{e:VEVD2} implies that $r_2>0$: the $(0,1)$-direction in the $(r_1,r_2)$-plane.

\paragraph{(ii)}
 Along ${\bf E}_{013}$, $X_0,X_1,X_{n+1},X_{n+2}=0$ but
 $(X_2,\cdots X_n)\neq(0,{\cdots},0)$, so $U(1)^2\to U_2(1)$. 
 Here,~\eqref{e:VEVD1} implies $r_1\geqslant 0$ while~\eqref{e:VEVD2} sets $r_2=0$: the $(1,0)$-direction. 
 When $m>2$, the (constrained) $X_{n+1},X_{n+2}\to0$ limit requires special care\cite{rBH-gB, Berglund:2022dgb}.

\paragraph{(iii)}
 Along ${\bf E}_{023}$, where $X_0,X_2,{\cdots},X_{n+2}=0$ but $X_1\,{\neq}\,0$,
 so $U(1)^2\to U_3(1)$. Then,~\eqref{e:VEVD1} sets $r_1=|x_1|^2\geqslant 0$ and substituting this into~\eqref{e:VEVD2} implies that $r_2=-m\,r_1$, which holds along the $(1,-m)$-direction.

\paragraph{(iv)}
 Along ${\bf E}_{123}$, where $(X_1,{\cdots},X_{n+2})=(0,{\cdots}0)$ but $X_0\,{\neq}\,0$, so $U(1)^2\to U_4(1)$. Then $r_1=-n|x_0|^2\leq 0$ from~\eqref{e:VEVD1}, and substituting this into~\eqref{e:VEVD2} implies that $r_2=-\frac{m{-}2}{n}\,r_1$, which holds along the $({-}n,m{-}2)$-direction.

The so-determined demarcation rays in the $(t_1,t_2)$-space are precisely as indicated by the (vertically read) $\skew0\vec{q}(X_*)$ 2-vectors~\eqref{e:q1-4} for the $X_*\neq0$ fields:
\begin{equation}
\begin{array}{@{}r|c|c|c|c@{}}
  &(\textit{i\/}) &(\textit{ii\/}) &(\textit{iii\/}) &(\textit{iv\/}) \\
 \textbf{nonzero} &X_{n+1},X_{n+2} &X_2,\cdots X_n &X_1 &X_0 \\ \toprule
 \pM{t_1\\t_2} &\pM{0\\[1pt]1} &\pM{1\\[1pt]0} &\pM{~~1\\-m} &\pM{-n\\m-2} \\ \bottomrule
\end{array}
\end{equation}
The preceding gauge $U(1)^2$-orbit partitioning thus {\em\/precisely\/} reproduces the vev-analysis\cite{rMP0, rBH-gB} and the phase-diagram in Figure~\ref{f:PhDiag}, also known as the Gelfand--Kapranov--Zelevinsky (GKZ) decomposition\cite{Oda:1991aa} and the ``secondary fan.''
\begin{figure}[htb]
$$
\vC{\TikZ{[very thick, scale=.75]
      \path[use as bounding box](-3,-3.3)--(4.2,2.3);
      \corner{(0,0)}{0}{90}{2}{green};
      \corner{(0,0)}{90}{{90+atan(3/1)}}{2}{red};
      \corner{(0,0)}{{90+atan(3/1)}}{{270+atan(1/3)}}{2}{blue};
      \corner{(0,0)}{{270+atan(1/3)}}{360}{2}{yellow};
      \draw[-stealth](0,0)--(1,0)node[above]{\scriptsize$(1,0)$};
       \path(2,0)node{(\textit{ii})};
      \draw[-stealth](0,0)--(0,1)node[right]{\scriptsize$(0,1)$};
       \path(0,1.8)node{(\textit{i})};
      \draw[-stealth](0,0)--(-3,1)node[above]{\scriptsize$(-n,m-2)$};
       \draw[thin, -stealth](-2,.667)to[out=-220, in=0]++(-.5,-.2)
        node[left=-2pt]{(\textit{iv})};
      \draw[-stealth](0,0)--(1,-3)node[above right]{\scriptsize$(1,-m)$};
       \draw[thin, -stealth](.667,-2)to[out=210, in=0]++(-.5,-.2)
        node[left=-2pt]{(\textit{iii})};
      \path(45:2)node{\Large\bf I};
      \path(-40:2)node{\Large\bf II};
      \path(225:2)node{\Large\bf III};
      \path(130:2)node{\Large\bf IV};
      \filldraw[fill=white](0,0)circle(1mm);
            }}
\parbox[c]{85mm}{\small\baselineskip=10pt\raggedright\noindent
$\bullet$~The ({\it i\/}) and ({\it ii\/}) demarcations are $m,n$-independent,
 and fix an integral basis/lattice in the $(t_1,t_2)$-plane\\[7pt]
$\bullet$~Only the ({\it iv}) demarcation, $(-n,m{-}2)$, depends on $n=\dim(\FF{m})$.\\[7pt]
$\bullet$~Only the ({\it iii\/}) and ({\it iv}) demarcations depend on the
 ``twist, '' $m$.\\[7pt]
$\bullet$~For $m{=}0$, the ({\it ii\/}) and ({\it iii\/}) demarcations coalesce,
 and so remove phase~II.
}
$$
 \caption{The ``phase diagram'' for the GLSM with the ambient space $\FF{m}$, shown here for $n=3$ and $m=3$.}
 \label{f:PhDiag}
\end{figure}
With the specified demarcations, Eqs.~\eqref{e:VEVD} determine the actual values of the nonzero vevs (omitting the usual $\langle{\cdots}\rangle$ notation and correcting\cite[Fig.~1]{rBH-gB}); see Table~\ref{t:PhValues}.
\begin{table}[htbp]
\caption{The vacuum expectation (background) values of the various fields in the various phases of the $\FF{m}$ GLSM.}
\centering
$\begin{array}{@{}r@{\:}|@{\:}c@{\:}|@{\:}c@{~}c@{~}c@{~}c@{\:}|@{\:}c@{}c@{}}
     &|X_0|&|X_1|&|X_2|&\cdots&|X_n|&|X_{n+1}|&|X_{n+2}| \\ \toprule
     \textbf{{\itshape i}\,}
                  & 0 & 0 & 0 &\cdots& 0 & * &  * \\
     \textbf{I\,}
                  & 0 & * & * &\cdots& * & * &  * \\
     \textbf{{\itshape ii}\,}
                  & 0 & 0 & * &\cdots& * & 0 &  0 \\
     \textbf{II\,}
                  & 0 & \text{see~\eqref{e:|x1|}} & * &\cdots& * & * &  * \\
     \textbf{{\itshape iii}\,}
                  & 0  & \sqrt{t_1}
                              & 0 &\cdots& 0 & 0 &  0 \\[1mm]
     \textbf{III\,}
                  & \sqrt{\frac{-mt_1{-}t_2}{(n{-}1)m{+}2}}
                        & \sqrt{\frac{(2{-}m)t_1{-}nt_2}{(n{-}1)m{+}2}}
                              & 0 &\cdots& 0 & 0 &  0 \\[1mm]
     \textbf{{\itshape iv}\,}
                  & \sqrt{-t_1/n}
                        & 0 & 0 &\cdots& 0 & 0 &  0 \\[1mm]
     \textbf{IV\,}
                  & \text{see~\eqref{e:|x1|}}
                        & 0 & 0 &\cdots& 0 & * &  * \\ \bottomrule
 \multicolumn8l{\text{``\,*\,'' denotes nonzero values; all vevs vanish at 
 $(t_1,t_2)\,\!=\!\,(0,0)$.}}
\end{array}$
\label{t:PhValues}
\end{table}%
\begin{equation}\textstyle
   |X_1|_{\rm II}=\sqrt{t_1-\sum_{i=2}^n|X_i|^2}~\overset!{\geq}0,\qquad
   |X_0|_{\rm IV}=\sqrt{\frac{|X_{n+1}|^2+|X_{n+2}|^2-t_1}{n}}.
 \label{e:|x1|}
\end{equation}
In particular, the vevs change continuously throughout the $(t_1,t_2)$-plane.

\begin{figure}[htb]
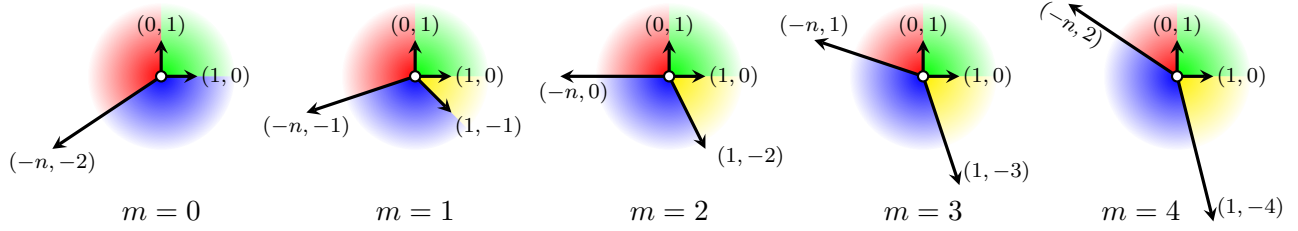

$$
\vC{\TikZ{[thick, scale=.48]
      \path[use as bounding box](-4,-3.5)--(31,2.3);
      \corner{(0,0)}{0}{90}{2}{green};
      \corner{(0,0)}{90}{{180+atan(2/3)}}{2}{red};
      \corner{(0,0)}{{180+atan(2/3)}}{360}{2}{blue};
      \draw[very thick, -stealth](0,0)--(1,0)node[right=-3pt]
        {\scriptsize$(1,0)$};
      \draw[very thick, -stealth](0,0)--(0,1)node[above=-3pt]
        {\scriptsize$(0,1)$};
      \draw[very thick, -stealth](0,0)--(-3,-2)node[below=-3pt]
        {\scriptsize$(-n,-2)$};
      \path(0,-3.7)node{$m=0$};
      \filldraw[fill=white](0,0)circle(4pt);
      \begin{scope}[xshift=7cm]
      \corner{(0,0)}{0}{90}{2}{green};
      \corner{(0,0)}{90}{{180+atan(1/3)}}{2}{red};
      \corner{(0,0)}{{180+atan(1/3)}}{315}{2}{blue};
      \corner{(0,0)}{315}{360}{2}{yellow};
      \draw[very thick, -stealth](0,0)--(1,0)node[right=-3pt]
        {\scriptsize$(1,0)$};
      \draw[very thick, -stealth](0,0)--(0,1)node[above=-3pt]
        {\scriptsize$(0,1)$};
      \draw[very thick, -stealth](0,0)--(-3,-1)node[below=-3pt]
        {\scriptsize$(-n,-1)$};
      \draw[very thick, -stealth](0,0)--(1,-1)node[below right=-3pt]
        {\scriptsize$(1,-1)$};
      \path(0,-3.7)node{$m=1$};
      \filldraw[fill=white](0,0)circle(4pt);
      \end{scope}
      \begin{scope}[xshift=14cm]
      \corner{(0,0)}{0}{90}{2}{green};
      \corner{(0,0)}{90}{180}{2}{red};
      \corner{(0,0)}{180}{{270+atan(1/2)}}{2}{blue};
      \corner{(0,0)}{{270+atan(1/2)}}{360}{2}{yellow};
      \draw[very thick, -stealth](0,0)--(1,0)node[right=-3pt]
        {\scriptsize$(1,0)$};
      \draw[very thick, -stealth](0,0)--(0,1)node[above=-3pt]
        {\scriptsize$(0,1)$};
      \draw[very thick, -stealth](0,0)--(-3,0)node[below=-2pt, xshift=5pt]
        {\scriptsize$(-n,0)$};
      \draw[very thick, -stealth](0,0)--(1,-2)node[right, yshift=-3pt]
        {\scriptsize$(1,-2)$};
      \path(0,-3.7)node{$m=2$};
      \filldraw[fill=white](0,0)circle(4pt);
      \end{scope}
      \begin{scope}[xshift=21cm]
      \corner{(0,0)}{0}{90}{2}{green};
      \corner{(0,0)}{90}{{90+atan(3/1)}}{2}{red};
      \corner{(0,0)}{{90+atan(3/1)}}{{270+atan(1/3)}}{2}{blue};
      \corner{(0,0)}{{270+atan(1/3)}}{360}{2}{yellow};
      \draw[very thick, -stealth](0,0)--(1,0)node[right=-3pt]
        {\scriptsize$(1,0)$};
      \draw[very thick, -stealth](0,0)--(0,1)node[above=-3pt]
        {\scriptsize$(0,1)$};
      \draw[very thick, -stealth](0,0)--(-3,1)node[above=-3pt]
        {\scriptsize$(-n,1)$};
      \draw[very thick, -stealth](0,0)--(1,-3)node[above right=-3pt]
        {\scriptsize$(1,-3)$};
      \path(0,-3.7)node{$m=3$};
      \filldraw[fill=white](0,0)circle(4pt);
      \end{scope}
      \begin{scope}[xshift=28cm]
      \corner{(0,0)}{0}{90}{2}{green};
      \corner{(0,0)}{90}{{90+atan(3/2)}}{2}{red};
      \corner{(0,0)}{{90+atan(3/2)}}{{270+atan(1/4)}}{2}{blue};
      \corner{(0,0)}{{270+atan(1/4)}}{360}{2}{yellow};
      \draw[very thick, -stealth](0,0)--(1,0)node[right=-3pt]
        {\scriptsize$(1,0)$};
      \draw[very thick, -stealth](0,0)--(0,1)node[above=-3pt]
        {\scriptsize$(0,1)$};
      \draw[very thick, -stealth](0,0)--(-3,2)node
        [below=-2pt, rotate=-30, xshift=6pt]{\scriptsize$(-n,2)$};
      \draw[very thick, -stealth](0,0)--(1,-4)node[above right=-3pt]
        {\scriptsize$(1,-4)$};
      \path(-1,-3.7)node{$m=4$};
      \filldraw[fill=white](0,0)circle(4pt);
      \end{scope}
            }}
$$
 \caption{The phase diagrams of the first three $\FF[3]m$-GLSMs. 
  Including $q_a(X_0)$, they also depict the anticanonical bundles, 
  $\mathcal{K}^*_{\smash{\FF{m}}}$, sections of which define 
  Ricci-flat hypersurfaces as their zero locus}
 \label{f:3F0-4}
\end{figure}

\paragraph{Aperiodicity}
The infinite sequence ($m=0,1,2,\dots$) of Hirzebruch surface scrolls, $\FF[2]m$, is well-know to provide only two diffeomorphism classes\cite{rH-Fm, rO-TV, rF-TV, rGrHa, rGE-CCAG, rCLS-TV, rKC-Fm}: the even-twisted $\FF[2]{2k}$ are all diffeomorphic to each other (``same, as smooth real manifolds''), as are the odd-twisted $\FF[2]{2k+1}$. We may write 
$\FF[2]{m}\approx_\IR\FF[2]{m\,(\mathrm{mod}\,2)}$, and find that analogously,
$\FF{m}\approx_\IR\FF{m\,(\mathrm{mod}\,n)}$\cite{rBH-gB}. In turn, the $\FF{m}$ for differing $m$ are all distinct as complex manifolds, and this is clearly seen from the phase diagrams in Figure~\ref{f:PhDiag}, reproduced in Figure~\ref{f:3F0-4} for $n=3$ and for $m=0,1,\cdots 4$. The sequence of Hirzebruch scrolls is in fact infinite, with the 
$\IP^{n-1}\!\hookrightarrow\!\FF{m}\!\twoheadrightarrow\!\IP^1$
bundle twist, $m$, unbounded. In fact, as the next section shows, there is an additional diversity for $n\geq3$ that grows with both $n$ and $m$.

The well-known $m$-dependent difference---{\em\/as complex manifolds\/}---of Hirzebruch {\em\/surface\/} scrolls, $\FF[2]m$, with different $m$-twists extends also to their higher-dimensional generalizations, $\FF{m}$\cite{rBH-gB}. Their distinguishing $m$-dependent characteristic is their unique {\em\/exceptional divisor\/} (complex codimension-1 irreducible submaniold), 
$\scS\!\subset_\IC\!\FF{m}$, with the maximally negative self-intersection:
$[\scS]^n={-}(n{-}1)m$. 
The Ricci-flat (Calabi--Yau) hypersurfaces 
$\scX\subset\FF{m}$ themselves being of complex codimen\-sion-1 therefore intersect this $\scS$:
\begin{equation}
  \dim\big(\scX\subset\FF{m}\big)=n{-}1, \quad\Rightarrow\quad
  \dim\big((\scX\cap\scS)\subset\FF{m}\big)=n{-}2,
\end{equation}
and so ``inherit'' distinguishing features stemming from $\scS\subset\FF{m}$ in codimension-2; those loci naturally turn up in
$H_{n-2}(\scX,\ZZ)\approx H^2(\scX,\ZZ)$ (both being dual to $H_2(\scX,\ZZ)$), and so provide a relevant distinction in stringy applications.

Already the first five phase diagrams in Figure~\ref{f:3F0-4} quite clearly do not exhibit the diffeomorphism $n$-cycle, 
$\FF{m}\approx_\IR\FF{m\,\mathrm{mod}\,n}$. 
Any putative $GL(2;\ZZ)$-transformation to map one to another is ruled out by the first $\bullet$-point in Figure~\ref{f:PhDiag}: the ({\em\/i\/})--({\em\/ii\/}) demarcations fix the basis, making the ({\em\/iii\/})--({\em\/iv\/}) demarcations unique for each $m$. This proves the inequivalence of the $\FF{m}$ GLSMs already at this semi-classical stage\cite{rPhases},
and the cumulative worldsheet instanton effects (following\cite{rMP0}) only compound the differences.

Notably, the phase diagrams are well-known to correspond not to the complex structure, but para\-metrize the (complexified) {\em\/K{\"a}hler structure\/}\cite{rPhases, rMP0} of the scrolls, and thus (properly restricted) also of the Ricci-flat hypersurfaces therein. 
Thus, the differences exhibited in Figure~\ref{f:3F0-4} actually pertain to (the GLSM dependence on) the K{\"a}hler structure of the Hirzebruch scrolls, their anticanonical bundles, and their Ricci-flat hypersurfaces!

\section{The Ambient-Space Deformation Family}
\label{s:AmbiDefo}
Calabi--Yau hypersurfaces in Hirzebruch scrolls, as studied recently\cite{rBH-gB, Berglund:2022dgb, Berglund:2024zuz, Hubsch:2025sph, Hubsch:2025teh, Hubsch:2026wxe}, benefit from the generalizing Hirzebruch's original construction\cite{rH-Fm}, by providing a concrete realization of the characteristic exceptional divisor, using the novel technique of generalized hypersurfaces and their complete intersections\cite{rgCICY1, rBH-Fm, rGG-gCI}, and then translating that into the complex-algebraic toric geometry framework; this provides two distinct but precisely correlated ways of computation. 

\subsection{Biprojective and Toric Renditions}
\label{s:2P-TV}
As mentioned above, the sequence of fiber bundle ambient spaces,
$\IP^{n-1}\!\hookrightarrow\!\FF{m}\!\twoheadrightarrow\!\IP^1$,
forms an unbounded-$m$ sequence, which furthermore harbors an additional diversity for $n\geq3$ that is growing combinatorially with $m$, and which we now discuss.
\paragraph{Biprojective Embedding.}
Following here\cite[Construction~2.1]{Berglund:2022dgb}, for every concrete hypersurface
\begin{subequations}
 \label{e:pe(x,y)}
\begin{alignat}9
  \FF{m;\epsilon} &:=
  \{p_{\epsilon\,}(x,y)=0\} \in \bgK[{r||c}{\IP^n&1\\\IP^1&m}],\\
  p_{\epsilon\,}(x,y) &:= 
     x_0y_0^m+x_1y_1^m +
     \sum_{i=0}^n\sum_{\ell=1}^{m-1} \epsilon_{i\ell}\,
                                     x_i\,y_0^{m-\ell}y_1^\ell,\quad
  \epsilon_{i\ell}\in\IC,
\end{alignat}
\end{subequations}
one can find a collection of mutually {\em\/algebraically independent\/}\footnote{In this context, $\frs_\alpha$ (defining $\scS_\alpha$) is ``algebraicaly independent'' from 
$\frs_\beta,\frs_\gamma$ means that no algebraic relationship 
$\frs_\alpha = ay_0^{b_0}y_1^{b_1}\frs_\beta +a'y_0^{c_0}y_1^{c_1}\frs_\gamma$ with $b_j,c_j\geq0$ and $a,a'\in\IC$ exists.} irreducible hypersurfaces,
$\big\{ \scS_\alpha := \{\frs_\alpha(x,y)$ $=\!0\}\subset\FF{m} \big\}$, 
each of maximally negative self-inter\-sec\-tion, and explicitly constructed as generalized hypersurfaces. This collection then also allows writing explicit deformation families of Calabi--Yau hypersurfaces, as well as their intersections with $\scS_\alpha$, to directly compute the corresponding ``inherited'' characteristics.

For example, Hirzebruch's original $p_0(x,y)=y_0^m+x_1y_1^m$ hypersurface has the following one exceptional divisor:
\begin{equation}
  \frs_0(x,y):=
  \bigg[ \Big(\frac{x_0}{y_1^m}-\frac{x_1}{y_0^m}\Big) 
              +\lambda \frac{p_0(x,y)}{(y_0y_1)^m} \bigg]
  =\bigg\{ \begin{array}{@{}l@{\quad}r@{\,}l@{\quad\text{when}~}r@{\,}l@{}}
          +2x_0/y_1^m, &\lambda&={+}1, &y_1&\neq0 ,\\[1mm]
          -2x_1/y_0^m, &\lambda&={-}1,&y_0&\neq0.\\
          \end{array}
 \label{e:s(x,y)}
\end{equation}
The key novelty here is that the defining holomorphic section, $\frs_0(x,y)$, is explicitly constructed as an equivalence class, where the two marginal representatives differ by
\begin{equation}
  \frs_0(x,y)\big|_{\lambda={+}1}-\frs_0(x,y)\big|_{\lambda={-}1}
  =2\frac{p_0(x,y)}{(y_0y_1)^m} \equiv 0~~\text{on all}~~\FF{m}.
 \label{e:2charts}
\end{equation}
That is, $\frs_0(x,y)$ is holomorphic on $\FF{m}$ although not on $\IP^n\!\times\!\IP^1$.

The underlying mechanism~\eqref{e:s(x,y)}--\eqref{e:2charts} insuring that 
$\frs_0(x,y)$ is in fact a holomorphic section is {\em\/precisely\/} the same one that enables the Wu-Yang construction of a magnetic monopole~\cite{rWY-MM}! There, physics can detect no difference between the different gauge potentials since they define the exactly same magnetic field. Here, the different 
($\lambda=\pm1$) representatives~\eqref{e:s(x,y)} specify parts of the same hypersurface, $\scS_0:=\{\frs_0(x,y)\,\!=\!\,0\}\subset\FF{m}$, over different charts in $\IP^1$, which agree perfectly on the overlap~\eqref{e:2charts} so the hypersurface is well-defined everywhere.

Since a Calabi--Yau hypersurface in $\FF{m}$ has to be of bi-degree $(n,2{-}m)$ in
$\IP^n\!\times\!\IP^1$, one considers
\begin{equation}
  f_0(x,y) \in 
  \Big(\!\oplus_{k=0}^2\big(c_k(x)\,y_0^{2-k}y_1^k\big)\!\Big)\,\frs_0(x,y),\quad
  \deg[c_k(x)]=n{-}1,
 \label{e:CYf(x,y)}
\end{equation}
and finds that~\eqref{e:CYf(x,y)} provides the full complement of requisite sections for $m\ge3$ when $c_1(\FF{m})$ is negative over the base-$\IP^1$, and the ``missing'' sections for the marginal case when $c_1(\FF{2})$ is null over the base-$\IP^1$\cite{rBH-gB}.
\begin{rem}\label{r:no1/dx}
To be holomorphic as guaranteed by~\eqref{e:s(x,y)}--\eqref{e:2charts}, the sections~\eqref{e:CYf(x,y)} may include only non-negative integral powers of the directrix.\hrulefill
\end{rem}

\paragraph{Toric Translation.}
The simple change of variables,
\begin{alignat}9
   (x_0,x_1,\cdots x_n;y_0,y_1) &\in  \IP^n\!\times\!\IP^1\\
   \to~~\nonumber
    \big( p_\epsilon(x,y), \frs_0(x,y)\!=\!X_1, \{x_i\!=\!X_i\}_{i=2,\cdots n},
          \{y_j\!=\!X_{n+j+1}\}_{j=0,1} \big) &\xrightarrow{p_\epsilon(x,y)=0}\FF{m},
 \label{e:2P-T}
\end{alignat}
provides a 1--1 ``translation'' from the bi-projective embedding~\eqref{e:pe(x,y)} to the toric specification~\eqref{e:q1-4}, which is well defined and a routine ``field redefinition'' in QFT, since its Jacobian is constant, the $(x;y)\mapsto(X)$ coordinate transformations moreover linear. Also, the $U(1;\IC)^2$-charges inherited directly from the $\IP^n\!\times\!\IP^1$ homogeneity degrees provide the toric $(\IC^*)^2$-action.

This ``translation'' extends in fact throughout the $\epsilon_{i\ell}$-deformation family~\eqref{e:pe(x,y)}\cite{Berglund:2022dgb}, and all directrices found as in Ref.\cite[Construction~2.1]{Berglund:2022dgb} are verified to be transverse, so each of their zero-locus is an irreducible divisor.
Suffice it here to provide a couple of examples, starting from $\FF[3]5$: 
\begin{subequations}
\begin{alignat}9
 && p_0(x,y)&= x_0\,y_0^5 +x_1\,y_1^5\quad\text{admits}\\
 \pM{~~1\\-5}{:}&\quad&
 \frs_{0}(x,y)
  &=\Big[\Big(\frac{x_0}{y_1^5} -\frac{x_1}{y_0^5}\Big)\Big/p_0(x,y)\Big];
\intertext{where ``$\big(\frac{x_0}{y_1^5}{-}\frac{x_1}{y_0^5}\big)/p_0(x,y)$'' abbreviates the precise definition in~\eqref{e:s(x,y)}. Then, the variable change~\eqref{e:2P-T} produces the toric rendition:}
 \FF[3]5:
  &&&\begin{array}{@{}ccccc}
              X_1 &X_2 &X_3 &X_4 &X_5 \\ \toprule
              1 &1   &1   &0   &0 \\[-2pt]
              -5  &0   &0   &1   &1 \\ \bottomrule
     \end{array}
 \label{e:3F5}
\end{alignat}
\end{subequations}
Next, consider two simple deformation terms (correcting\cite{Berglund:2022dgb}):
\begin{subequations}
\begin{alignat}9
 \label{e:p41}
 && p_1(x,y)&= x_0\,y_0^5 +x_1\,y_1^5 +x_2\,y_0^4 y_1^1\quad\text{admits}\\
 \pM{~~1\\-4}{:}&\quad&
 \frs_{11}(x,y)
         &=\Big[\Big( \frac{x_0\,y_0}{y_1^5} 
                      -\frac{x_1}{y_0^4}
                      +\frac{x_2}{y_1^4} \Big)\Big/p_1(x,y) \Big],\\
 \pM{~~1\\-1}{:}&\quad&
 \frs_{12}(x,y)
         &=\Big[\Big( \frac{x_0}{y_1}
                      -\frac{x_1\,y_1^4}{y_0^5}
                      -\frac{x_2}{y_0} \Big)\Big/p_1(x,y) \Big];
\intertext{where the constant-Jacobian $(x_0,x_1,x_2,x_3;y_0,y_1)\to(p_1,\frs_{11},\frs_{12},x_3;y_,y_1)$ change of variables and after setting $p_1=0$ now produces:}
 \FF[3]{\!\sss(4,1,0)}:
 &&&\begin{array}{@{}ccccc}
              X_1 &X_2 &X_3 &X_4 &X_5 \\ \toprule
              1   &1   &1   &0   &0 \\[-2pt]
              -4  &-1  &0   &1   &1 \\ \bottomrule
    \end{array}
\end{alignat}
\end{subequations}
and also:
\begin{subequations}
\begin{alignat}9
 \label{e:p32}
 && p_2(x,y)&= x_0\,y_0^5 +x_1\,y_1^5 +x_2\,y_0^3\,y_1^2\quad\text{admits}\\
 \label{e:p32d1}
 \pM{~~1\\-3}{:}&\quad&
 \frs_{21}(x,y)
         &=\Big[\Big( \frac{x_0\,y_0^2}{y_1^5} 
                      -\frac{x_1}{y_0^3} 
                      +\frac{x_2}{y_1^3} \Big)\Big/p_2(x,y) \Big],\\
 \label{e:p32d2}
 \pM{~~1\\-2}{:}&\quad&
 \frs_{22}(x,y)
         &=\Big[\Big( \frac{x_0}{y_1^2} 
                      -\frac{x_1 y_1^3}{y_0^5} 
                      -\frac{x_2}{y_0^2} \Big)\Big/p_2(x,y) \Big];\\
 \FF[3]{\!\sss(3,2,0)}:
 &&&\begin{array}{@{}ccccc}
              X_1 &X_2 &X_3 &X_4 &X_5 \\ \toprule
              1   &1   &1   &0   &0 \\[-2pt]
              -3  &-2  &0   &1   &1 \\ \bottomrule
    \end{array}
\end{alignat}
\end{subequations}
In both cases, the identification $X_1\!=\!\frs_{11}(x,y)$ and $X_2\!=\!\frs_{12}(x,y)$, i.e., $X_1\!=\!\frs_{21}(x,y)$ and $X_2\!=\!\frs_{22}(x,y)$ is evident from the degrees. In both cases, the deg-$\pM{~~1\\-5}$ directrix has split into two, the $\IP^1$-degrees of which add up: $-4-1=-5$ as well as $-3-2=-5$.

Combining the two deformations predictably ``distributes'' the system of directrices further: it modifies the deg-$\pM{~~1\\-3}$ directrix and splits the deg-$\pM{~~1\\-2}$ directrix into two {\em\/distinct\/} deg-$\pM{~~1\\-1}$ directrices:
\begin{subequations}
\begin{alignat}9
 \label{e:p311}
 && p_3(x,y)&= x_0\,y_0^5 +x_1\,y_1^5 +x_2\,y_0^4 y_1^1 +x_3\,y_0^3\,y_1\!^2
    \quad\text{admits}\\
 \label{e:p311d1}
 \pM{~~1\\-3}{:}&\quad&
 \frs_{31}(x,y)
         &=\Big[\Big( \frac{x_0\,y_0^2}{y_1^5}  
                      -\frac{x_1}{y_0^3}
                      +\frac{x_2y_0}{y_1^4} 
                      +\frac{x_3}{y_1^3} \Big)\Big/p_3(x,y) \Big],\\
 \label{e:p311d2}
 \pM{~~1\\-1}{:}&\quad&
 \frs_{32}(x,y)
         &=\Big[\Big( \frac{x_0}{y_1} 
                      -\frac{x_1 y_1^4}{y_0^5} 
                      -\frac{x_2}{y_0}
                      -\frac{x_3 y_1}{y_0^2} \Big)\Big/p_3(x,y) \Big],\\
 \label{e:p311d3}
 \pM{~~1\\-1}{:}&\quad&
 \frs_{33}(x,y)
         &=\Big[\Big( \frac{x_0 y_0}{y_1^2} 
                      -\frac{x_1 y_1^3}{y_0^4} 
                      +\frac{x_2}{y_1} 
                      -\frac{x_3}{y_0}\Big)\Big/p_3(x,y) \Big];\\
 \FF[3]{\!\sss(3,1,1)}:
 &&&\begin{array}{@{}ccccc}
              X_1 &X_2 &X_3 &X_4 &X_5 \\ \toprule
              1   &1   &1   &0   &0 \\[-2pt]
              -3  &-1  &-1  &1   &1 \\ \bottomrule
    \end{array}
 \label{e:311}
\end{alignat}
\end{subequations}
Although both have the same deg-$\pM{~~1\\-1}$, $X_2=\frs_{32}$ and $X_3=\frs_{33}$ are algebraically independent, and the corresponding \eqref{e:2P-T}-like ``bipolar-to-toric translation'' variable changes each have a constant Jacobian. 

\subsection{Twist-Reduction}
More notably, a novel feature emerges in~\eqref{e:311}: The fact that no coordinate in~\eqref{e:311} is invariant with respect to $q_2$ indicates that the gauge $U(1;\IC)^2$-symmetry of the corresponding GLSM may be adjusted, defining $\widetilde{q\,}\!_2=q_2{+}q_1$, which results in:
\begin{equation}
  \FF[3]{\!\sss(3,1,1)}:~
  \begin{array}{@{}c@{~}c@{~}c@{~}c@{~}c@{}}
              X_1 &X_2 &X_3 &X_4 &X_5 \\ \toprule
              1   &1   &1   &0   &0 \\[-2pt]
              -3  &-1  &-1  &1   &1 \\ \bottomrule
  \end{array}
  ~~\xrightarrow{\sss q_2{+}q_1=\skew3\widetilde{q}_2}~~
  \begin{array}{@{}c@{~}c@{~}c@{~}c@{~}c@{}}
              X_1 &X_2 &X_3 &X_4 &X_5 \\ \toprule
              1   &1   &1   &0   &0 \\[-2pt]
              -2  &0   &0   &1   &1 \\ \bottomrule
  \end{array}
  ~:\FF[3]{\!\sss(2,0,0)}\!=\!\FF[3]2,
 \label{e:311-200}
\end{equation}
and which reminds of the expected {\em\/diffeomorphism,\/} 
$\FF[3]5\approx_\IR\FF[3]{\smash{5\,(\mathrm{mod}\,3)}}=\FF[3]2$, for which of course no holomorphic coordinate-level mapping can exist as there is no such complex bijection.
\begin{rem}\label{r:nFlp}
The change of $U(1)^2$/$q_a$-basis~\eqref{e:311-200} is \underline{\em\/neither\/} a so-called {\em\/flip\/} nor a {\em\/flop\/}\cite{rCLS-TV}! Its precise algebro-geometric explanation remains to be determined, and the preceding explicit complex coordinate-level rational variable changes (one would hope) should help making this determination.
Furthermore, since $q_2=\skew3\widetilde{q}_2{-}q_1$ is {\em\/not\/} a 
non-negative linear combination of $q_1,\skew3\widetilde{q}_2$, the perfectly routine $U(1)^2$ basis-change~\eqref{e:311-200} is not the identification of the standard Mori vectors, nor could $q_2(X_i)$ serve as a Mori vector since none of its components vanish.\hrulefill
\end{rem}
Nevertheless, there {\em\/do exist\/} complex coordinate-level albeit {\em\/rational\/} variable changes that do map the biprojective rendition~\eqref{e:p311} of $\FF[3]{\!\sss(3,1,1)}$ to the biprojective embedding of $\FF[3]2$! For example:
\begin{subequations}
 \label{e:311-2}
\begin{alignat}9
 (x_0, x_1, x_2, x_3; y_0, y_1 )
  &\mapsto
   \big( \tfrac{\xi_0}{\eta_0^3},
         \tfrac{\xi_1}{\eta_1^3},
         \tfrac{\xi_2}{\eta_0^2\eta_1},
         \tfrac{\xi_3}{\eta_0\eta_1^2};
         \eta_0,
         \eta_1 \big)\\*
 x_0 y_0^5 {+}x_1 y_1^5 {+}x_2 y_0^4 y_1^1 {+}x_3 y_0^3 y_1^2
  &\to (\xi_0{+}\xi_2{+}\xi_3)\eta_0^2 +\xi_1\eta_1^2.
 \label{e:311-2a}
\end{alignat}
Similarly, the slightly modified variable change produces:
\begin{alignat}9
 (x_0, x_1, x_2, x_3; y_0, y_1 )
  &\mapsto
    \big( \tfrac{\xi_0}{\eta_0^3},
          \tfrac{\xi_1}{\eta_1^3},
          \tfrac{\xi_2}{\eta_0^2\eta_1},
          \tfrac{\xi_3}{\eta_0^2\eta_1};
          \eta_0,
          \eta_1 \big)\\*
 x_0 y_0^5 {+}x_1 y_1^5 {+}x_2 y_0^4 y_1^1 {+}x_3 y_0^3 y_1^2
  &\to (\xi_0{+}\xi_2)\eta_0^2{+}\xi_1\eta_1^2{+}\xi_3\eta_1\eta_0.
 \label{e:311-2b}
\end{alignat}
Both~\eqref{e:311-2a} and~\eqref{e:311-2b} result in Hirzebruch-like
deg-$\pM{1\\2}$ defining equations of $\FF[3]2\subset\IP^2\!\times\!\IP^1$, and so are equivalent by {\em\/linear\/} reparametrization. Indeed, an aprorpiate linear $\xi$-variable redefinition results in the simple mapping:
\begin{alignat}9
 \label{e:311-200f}
 (x_0, x_1, x_2, x_3; y_0, y_1 )
  &\mapsto \big( \tfrac{\xi_0}{\eta_0^3}{-}\xi_2\eta_0^2\eta_1,
                 \tfrac{\xi_1}{\eta_1^3}{-}\xi_3\eta_0^3,
                 \xi_2\eta_0^3,
                 \xi_3\eta_1^3;
                 \eta_0, 
                 \eta_1 \big)\\*
 x_0 y_0^5 {+}x_1 y_1^5 {+}x_2 y_0^4 y_1^1 {+}x_3 y_0^3 y_1^2
  &\to \xi_0\eta_0^2 +\xi_1\eta_1^2 ~:\FF[3]2=\ssK[{r||c}{\IP^3&1\\ \IP^1&2}].
\end{alignat}
\end{subequations}
The variable change~\eqref{e:311-200f} not only produces precisely the Hirzebruch-canonical biprojective embedding of $\FF[3]2\subset\IP^2\!\times\!\IP^1$, but even has a constant Jacobian! In turn, neither of the directrices~\eqref{e:p311d1}--\eqref{e:p311d3} ``survive'': none of them satisfy the conditions detailed in Ref.\cite[Construction~2.1]{Berglund:2022dgb}, are not holomorphic as specified in~\eqref{e:s(x,y)}--\eqref{e:2charts}, and so do not define any irreducible submanifold. 
However, it is straightforward to find the \eqref{e:s(x,y)}-like directrix,
\begin{equation}
  \Big\{
  \frs_0':=\Big[\Big(\! \frac{\xi_0}{\eta_1^2}-\frac{\xi_1}{\eta_0^2} \!\Big)
                 \Big/\big(\xi_0\eta_0^2{+}\xi_1\eta_1^2\big) \Big]
  =0\Big\} = \scS_0 \subset \FF[3]2=\ssK[{r||c}{\IP^3&1\\ \IP^1&2}].
\end{equation}

In fact, similar constant-Jacobian (albeit rational) variable changes can be found throughout the $\epsilon_{i\ell}$-deformation family when the $q_2(X_i)$ become all nonzero such as~\eqref{e:311-200}. 

Comparing the deformations:
\begin{alignat}9
  \eqref{e:p41}\!:~~ &p_0(x,y)+x_2\,y_1^4 y_0^1\quad
  &\leadsto&~\FF[3]{\!\sss(4,1,0)},\\
  \eqref{e:p32}\!:~~ &p_0(x,y)+x_2\,y_1^3 y_0^2\quad 
  &\leadsto&~\FF[3]{\!\sss(3,2,0)},\\
  \eqref{e:p311}\!:~~&p_0(x,y)+x_2\,y_1^4 y_0^1 +x_2\,y_1^3 y_0^2\quad
  &\leadsto&~\FF[3]{\!\sss(3,1,1)},
\end{alignat}
it may be tempting to hazard a guess that the exponents in the deformation directly imply the ``distribution'' of the twist,
$m=5\leadsto(4,1)$, $\leadsto(3,2)$, and $\leadsto(3,1,1)$.

However, that this correlation is certainly more complicated is demonstrated by considering the {\em\/symmetric\/} combined deformation:
\begin{subequations}
\begin{alignat}9
 \label{e:p311'}
 && p_4(x,y)&= x_0\,y_0^5 +x_1\,y_1^5 +x_2\,y_0^4 y_1^1 +x_3\,y_0^1\,y_1^4
    \quad\text{admits}\\
 \pM{~~1\\-3}{:}&\quad&
 \frs_{41}(x,y)
         &=\Big[\Big( \frac{x_0 y_0}{y_1^4}
                      -\frac{x_1 y_1}{y_0^4}
                      +\frac{x_2}{y_1^3}
                      -\frac{x_3}{y_0^3} \Big)\Big/p_4(x,y) \Big],
 \label{e:p311'd1}\\
 \pM{~~1\\-1}{:}&\quad&
 \frs_{42}(x,y)
         &=\Big[\Big( \frac{x_0}{y_1}
                      -\frac{x_1 y_1^4}{y_0^5}
                      -\frac{x_2}{y_0}
                      -\frac{x_3 y_1^3}{y_0^4} \Big)\Big/p_4(x,y) \Big],
 \label{e:p311'd2}\\
 \pM{~~1\\-1}{:}&\quad&
 \frs_{43}(x,y)
         &=\Big[\Big( \frac{x_0 y_0^4}{y_1^5}
                      -\frac{x_1}{y_0}
                      +\frac{x_2 y_0^3}{y_1^4}
                      +\frac{x_3}{y_1} \Big)\Big/p_4(x,y) \Big];
 \label{e:p311'd3}\\
 \FF[3]{\!\sss(3,1,1)'}:
 &&&\begin{array}{@{}ccccc}
              X_1 &X_2 &X_3 &X_4 &X_5 \\ \toprule
              1   &1   &1   &0   &0 \\[-2pt]
              -3  &-1  &-1  &1   &1 \\ \bottomrule
    \end{array}
 \label{e:311'}
\end{alignat}
\end{subequations}
Although the toric rendition~\eqref{e:311'} seems indistinguishable from~\eqref{e:311}, their biprojective embeddings differ, as does the collection of hallmark directrices. Ultimately for GLSM and superstring application, what matters is whether there exist Calabi--Yau hypersurfaces in one that cannot be found in the other, but we defer that question to a separate effort.

The complex coordinate-level, the constant-Jacobian rational mapping 
$\FF[3]{\!\sss(3,1,1)'}\to\FF[3]2$ here turns out to differ only slightly from~\eqref{e:311-2}:
\begin{subequations}
 \label{e:311'-2}
\begin{small}
\begin{alignat}9
 \label{e:311'-200f}
 \FF[3]{\!\sss(3,1,1)'}\,{:}~(x_0, x_1, x_2, x_3, y_0, y_1 )
  &\mapsto \big( \tfrac{\xi_0}{\eta_0^3}{-}\xi_2\eta_0^2\eta_1,
                 \tfrac{\xi_1}{\eta_1^3}{-}\xi_3\eta_0\eta_1^2,
                 \xi_2\eta_0^3,
                 \xi_3\eta_1^3,
                 \eta_0, 
                 \eta_1 \big)\\*
 x_0 y_0^5 {+}x_1 y_1^5 {+}x_2 y_0^4 y_1^1 {+}x_3 y_0^1 y_1^4
  &\to \xi_0\eta_0^2 +\xi_1\eta_1^2 ~{:}\,\FF[3]2\!=\!\bgK[{r||c}{\IP^3&1\\ \IP^1&2}].
\end{alignat}
\end{small}%
\end{subequations}
Similarly, while~\eqref{e:p311} {\em\/can\/} be turned into~\eqref{e:p311'} by means of the (also rational) variable change,
$x_3\to x_3\big(\frac{y_1}{y_0}\big)^2$, the Jacobian of this change is
$\big(\frac{y_1}{y_0}\big)^2$---ill defined at the poles of $\IP^1_y$. Also, this change of variables {\em\/does not\/} map the hallmark triple of directrices~\eqref{e:p311d1}--\eqref{e:p311d3} 
into~\eqref{e:p311'd1}--\eqref{e:p311'd3}!
\begin{rem}\label{r:nTV}
This disagreement in directrices then suggests that the hypersurfaces~\eqref{e:p311} and~\eqref{e:p311'} may well, in fact, be {\em\/two distinct\/} complex manifolds, both corresponding to the toric variety $\FF[3]{\!\sss(3,1,1)}$ but differing in some subtle way that does not seem to be detectable by standard (complex-algebraic) toric geometry.\hrulefill
\end{rem}
While the precise algebro-geometric nature of the constant-Jacobian complex coordinate-level rational mappings such as~\eqref{e:311-2} and~\eqref{e:311'-2} remains to be determined, it is clear that these are rather straightforward field redefinitions of the kind that are standard in QFT in general, and so also for GLSMs. With this proviso, the foregoing discussion makes it rather clear:
\begin{cor}\label{c:defFm}
The biprojective embedding of Hirzebruch surfaces $\FF{m}$ form an explicit deformation family where the twist-degree (\/$m$) and the hallmark directrix in the ``central'' model, $\FF{m}$, ``distribute'' with deformation.
Deformations in the family that are increasingly more generic reach models that are (via rational coordinate change) less-twisted, all the way to the lowest-twisted model within the diffeomorphism class, $\FF{m\,(\mathrm{mod}\,n)}$.
\end{cor}
\begin{figure}[htb]
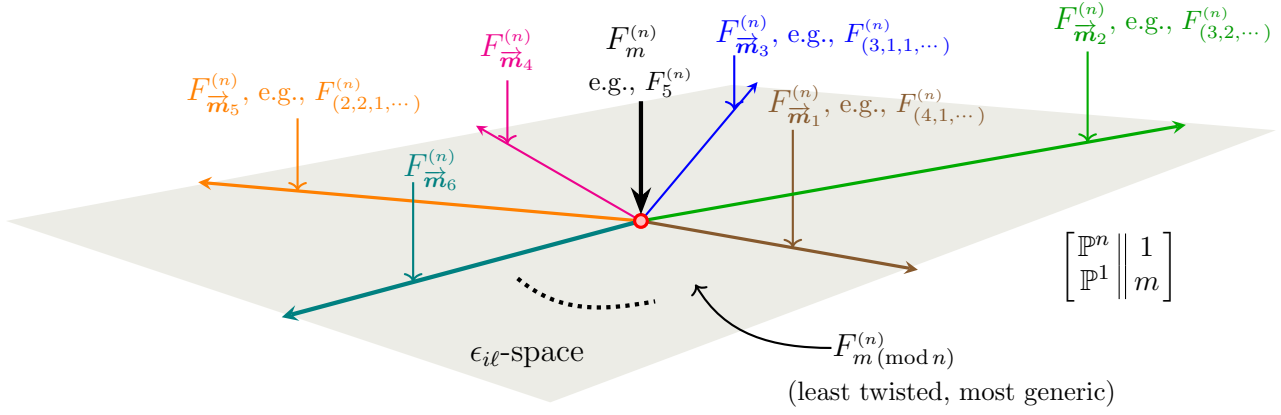

 \begin{center}
  \TikZ{[scale=1.2]\path[use as bounding box](-7,-1.8)--(8.2,2.0);
        \fill[yellow!20!gray!20](-7,0)--(-1,-2)--(7,1)--(1,1.5);
         \path(-2,-1.5)node[right]{\large$\epsilon_{i\ell}$-space};
        \draw[brown!70!black,very thick,-stealth](0,0)--++(-10:3.1);
         \path[brown!70!black](-10:1.3)+(0,1.5)node[right]
              {{\large$\FF{\overrightarrow{\boldsymbol{m}}_1}$}\small, e.g., 
                $\FF{(4,1,\cdots)}$};
         \draw[brown!70!black,<-,thick](-10:1.7)--++(0,1.3);
        \draw[very thick,green!67!black,-stealth](0,0)--++(10:6.1);
         \path[green!60!black](10:4.5)+(0,1.4)node[right]
              {{\large$\FF{\overrightarrow{\boldsymbol{m}}_2}$}\small, e.g., 
                $\FF{(3,2,\cdots)}$};
         \draw[green!67!black,<-,thick](10:5)--++(0,1);
        \draw[thick,blue,-stealth](0,0)--++(50:2);
         \path[blue](50:1.1)+(0,1.2)node[right]
              {{\large$\FF{\overrightarrow{\boldsymbol{m}}_3}$}\small, e.g., 
                $\FF{(3,1,1,\cdots)}$};
         \draw[blue,<-,thick](50:1.6)--++(0,.6);
        \draw[thick,magenta,-stealth](0,0)--++(150:2.1);
         \path[magenta](150:1.7)+(0,.7)node[above]
              {\large$\FF{\overrightarrow{\boldsymbol{m}}_4}$};
         \draw[magenta,<-,thick](150:1.7)--++(0,.7);
        \draw[very thick,orange,-stealth](0,0)--++(175:4.9);
         \path[orange](175:2.3)+(0,1.2)node[left]
              {{\large$\FF{\overrightarrow{\boldsymbol{m}}_5}$}\small, e.g., 
                $\FF{(2,2,1,\cdots)}$};
         \draw[orange,<-,thick](175:3.8)--++(0,.8);
        \draw[ultra thick,teal,-stealth](0,0)--++(195:4.1);
         \path[teal](195:2.6)+(.2,.9)node[above]
              {\large$\FF{\overrightarrow{\boldsymbol{m}}_6}$};
         \draw[teal,<-,thick](195:2.6)--++(0,1.1);
        \draw[ultra thick, dotted](205:1.5)to[out=-40,in=190](280:.9);
        \filldraw[red,fill=pink,very thick](0,0)circle(.7mm);
         \draw[Stealth-, ultra thick](0,.07)--++(0,1.25);
         \path(0,1.25)node[above]{\small e.g., $\FF5$};
         \path(-.1,1.7)node[above]{\large$\FF{m}$};
         \draw[<-,thick](.6,-.7)to[out=-60,in=180]++(1.5,-.7);
        \path(2,-1.4)node[right]{\large$\FF{\smash{m\,(\mathrm{mod}\,n)}}$};
        \path(1.5,-1.9)node[right]{\small(least twisted, most generic)};
        \path(4.5,-.5)node[right]{$\bgK[{r||c}{\IP^n&1\\ \IP^1&m}]$};
        }
 \end{center}
 \caption{A `big-picture' sketch of the full deformation family of degree-$\pM{1\\m}$ hypersurfaces in $\IP^n\times\IP^1$, i.e., distinct Hirzebruch scrolls.}
 \label{f:disDef}
\end{figure}
The situation is illustrated in Figure~\ref{f:disDef}, which depicts the $\epsilon_{i\ell}$-deformation family containing all Hirzebruch scrolls that are diffeomorphic to the central $\FF{m}$ and have twists with (taxicab length of the) twist no more than $m$: $\overrightarrow{m}=(m_1,\cdots m_n)$ with $m_i\geq0$ and $\sum_i m_i\geq m$. Since every $\FF{\sss\overrightarrow{m}}$ is a $\overrightarrow{m}$-twisted $\IP^{n-1}$-bundle over $\IP^1$, the toric realization of $\FF{\sss\overrightarrow{m}}$ reveals that the twisting is defined relative to the fan of the fiber-$\IP^{n-1}$---which exhibits a dihedral symmetry, $D_n$. This yields a provisional classification of toric varieties to be found within the explicit deformation family of biprojective embeddings,
$\FF{m}\in\ssK[{r||c}{\IP^n&1\\ \IP^1&m}]$; see Ref.~\cite{Hubsch:2026wxe} for more details.

This then provides for the unbounded collection of the ``well-known'' ambient spaces, $A=\FF{m}$, in which to embed the Ricci-flat/Calabi--Yau hypersurfaces of ultimate interest.

\section{Calabi--Yau Hypersurfaces in Hirzebruch Scrolls}
Having organized the current collection of ``well-known'' ambient spaces, 
$\FF{m}$, we may now turn to the actual interest: embedding the Ricci-flat/Calabi--Yau target spaces as hypersurfaces in each of the $\FF{m}$ in the $\epsilon_{i\ell}$-family such as depicted in Figure~\ref{f:disDef}. To this end, we need anticanonical sections from which to choose the hypersurface-defining constraint equation and GLSM superpotential.

\subsection{Laurent Hypersurfaces}
\label{s:DefoPic}
For the cases $m=0,1,2$, the constructions are well known, specified through-and-through within the standard framework of complex algebraic toric geometry\cite{rGrHa, rKKMS-TE1, rD-TV, rO-TV, rF-TV, rGE-CCAG, rCLS-TV, rCK}, and in this sense harbor no novelty; see Ref.\cite{rBH-Fm} for computational details.

\paragraph{Unsmoothable Tyurin Degeneration.}
For $m\ge3$, all deg-$\pM{n\\2-m}$ defining sections, $f(x,y)$ on $\FF{m;0}$, as defined in~\eqref{e:CYf(x,y)} and~\eqref{e:pe(x,y)}, respectively, {\em\/factorize.} Consequently, all such Calabi--Yau hypersurfaces reduce to a union of two ``lobes,''
\begin{equation}
  \{f_0(x,y)\!=\!0\} =
  \Big\{ \underbrace{\Big(\!\textstyle\sum_{k=0}^2\big(c_k(x)\,y_0^{2-k}y_1^k\big)
                      \!\Big) =0}_{\mathscr{C}}\Big\} \cup
  \Big\{ \underbrace{ \frs_0(x,y) =0}_{\mathscr{D}} \Big\}.
 \label{e:Xsh}
\end{equation}
The latter of these, $\mathscr{D}$, is called the {\em\/directrix\/}\cite{rGrHa}, so the comlement, $\mathscr{C}$, was termed the {\em\/complementrix\/}\cite{Berglund:2022dgb}. Even for the most general choices of $\mathscr{C}$, the Calabi--Yau hypersurface~\eqref{e:Xsh} is ``Tyurin degenerate''\cite{tyurin2003fano}, and singular at the intersection
\begin{equation}
  \mathrm{Sing}\big[\{f_0(x,y)\!=\!0\}\big] =
  \mathscr{C}\cap\mathscr{D}
  \in\bgK[{r||ccc}{\IP^n&1&n-1&~~1\\ \IP^1&m&2&-m}],
 \label{e:shX}
\end{equation}
which is itself a Calabi--Yau $(n{-}1)$-fold within the Calabi--Yau $n$-fold~\eqref{e:Xsh}, a {\em\/matriyoshka.} 

Given the list of ``coordinate fields'' and their charges in the toric rendition~\eqref{e:3F5}, it is straightforward to list all {\em\/regular\/} 
deg-$\pM{~~3\\-3}$ polynomials (regular sections of the anticanonical bundle, with non-negative exponents for each $X_i$), to satisfy the anomaly cancellation condition~\eqref{e:CY} and define Ricci-flat, i.e., Calabi--Yau hypersurface ground states (omitting ``$\otimes$'' between factors, for brevity of display):
\begin{equation}
  X_1^3(X_4{\oplus}X_5)^{12} \oplus 
  X_1^2(X_2{\oplus}X_3)(X_4{\oplus}X_5)^7 \oplus 
  X_1(X_2{\oplus}X_3)^2(X_4{\oplus}X_5)^2.
 \label{e:K*s}
\end{equation}
Evidently, every polynomial of this form has an $X_1=\frs_0(x,y)$ factor, i.e., factorizes as described in~\eqref{e:Xsh}--\eqref{e:shX}. Since there exist no $X_1$-independent deg-$\pM{~~3\\-3}$ monomials, zero loci of~\eqref{e:Xsh}, i.e.,~\eqref{e:K*s} are deemed \underline{\em\/unsmoothable\/} in standard complex-algebraic toric geometry. Indeed, the same happens with all Hirzebruch scrolls for $m\geq3$: the \eqref{e:CY}-``requisite degree'' of anticanonical sections, equal to the $U(1;\IC)^2$-charges and proportional to the 1st Chern class, is negative in the 2nd component; $c_1(\FF{m})\not\geq0$.

The defining sections~\eqref{e:Xsh} are parametrized by the choice of the polynomials $c_k(x)$, which defines an explicit deformation family of Calabi--Yau hypersurfaces in $\FF{m}$; this family is associated with the central point in the Figure~\ref{f:disDef}.
Much the same, Calabi--Yau hypersurfaces in $\epsilon_{i\ell}$-deformed Hirzebruch surfaces and analogously parametrized by some $c_k(x)$-like polynomials, and their explicit deformation families are then associated with the corresponding points in the $\epsilon_{i\ell}$-plane, thus defining a fibration of 
the deformation families of Calabi--Yau hypersurfaces in Hirzebruch scrolls over 
the deformation families Hirzebruch scrolls with bounded (taxicab length of the) total twist, $|m|=\sum_i m_i\leq m$.

\paragraph{Laurent Smoothing.}
Since all regular monomials~\eqref{e:K*s} vanish at the singular locus~\eqref{e:shX}, i.e., $\{X_1=0\}$, Refs.\cite{rBH-gB, Berglund:2022dgb} explored smoothing the singularity of~\eqref{e:Xsh} by deforming the hypersurface away from the singular locus~\eqref{e:shX}---by the only way possible: by including {\em\/rational\/} monomials,
\begin{equation}
  (X_2{\oplus}X_3)^3\big( \oplus_{k=0}^3 X_4^{k-3} X_5^{-k} \big)
 \label{e:K*sr}
\end{equation} which are {\em\/not\/} of the general form~\eqref{e:s(x,y)}, but will be motivated in a deformation-theoretic way below. Adding a nonzero multiple of even just one such monomial to an otherwise general but regular polynomial~\eqref{e:K*s} renders the resulting Laurent polynomial {\em\/transverse\/}:
\begin{equation}
  \{ f(X)=0=\rd f(X) \} \subset
  {\bf E}_{12}\cup{\bf E}_3,~ {\bf E}_1\cup{\bf E}_{23},
\end{equation}
which are {\em\/excised\/} from the ``geometric'' phases~\eqref{e:nFmQuot} and~\eqref{e:buWCP}, where they define smooth Calabi--Yau hypersurfaces,
$\scX_5^{\sss(2)}\in\FF[3]5[c_1]$, in fact K3 surfaces. 

The putative poles, stemming from the vanishing of $X_4,X_5$ in the denominators~\eqref{e:K*sr}, require the use of an ``intrinsic limit'' akin to L\!'Hopital's rule\cite{Berglund:2022dgb}---which goes firmly beyond the standard framework of complex-algebraic geometry. In turn, the use of such constrained limits is rather standard in QFT and physics in general, and so is not outside the framework of GLSM, just not needed until recently.

\subsection{Deformation Motivations}
The key condition~\eqref{e:CY} always has an obvious solution, the so-called ``fundamental monomial''\cite{rHY-SL2}, $\Pi X\,{:=}\,\prod_iX_i$, so that the ``defining function,'' $f(X)$, in the superpotential, $X_0\,f(X)$, may be regarded as a deformation of $\Pi X$.

\paragraph{The Building Blocks.}
Indeed, all by itself, the hypersurface defined by $\Pi X=0$ is highly singular, albeit combinatorially very simple:\footnote{I wish to thank Fedya Bogomolov for drawing my attention to this locus in 1991; I can only bemoan not having paid more attention and heeding his recommendations $\sfrac{1\!}3$ of a century earlier.} It is the locus where precisely one of all the $X_i$ vanishes at a time, which then together form the union of $X_i$-hyperplanes,
$\bigcup_i\{X_i=0\}\subset\FF{m}$. 
This union has an entire hierarchy of singularities: at
the codimension-1 intersection of each two such hyperplanes, which are themselves singular at
the codimension-2 common intersection of three such hyperplanes, and so on, down to the codimension-$(n{-}1)$ singular points---the vertices of this
{\em\/polyhoron.} (The multiple intersections are limited by the structure of the Stanley-Reisner ideal\cite{rCLS-TV}, so this polyhoron is not the regular, simplex-shaped polyhoron at ``infinity'' of the so-called Dwork pencil of $\IP^n[n{+}1]$ hypersurfaces\cite{Dwork:1969p-a}, which started the exploration of mirror symmetry\cite{Candelas:1990qd, Candelas:1990rm, Morrison:1991cd, rPeriods1}.)

\begin{figure}[htbp]
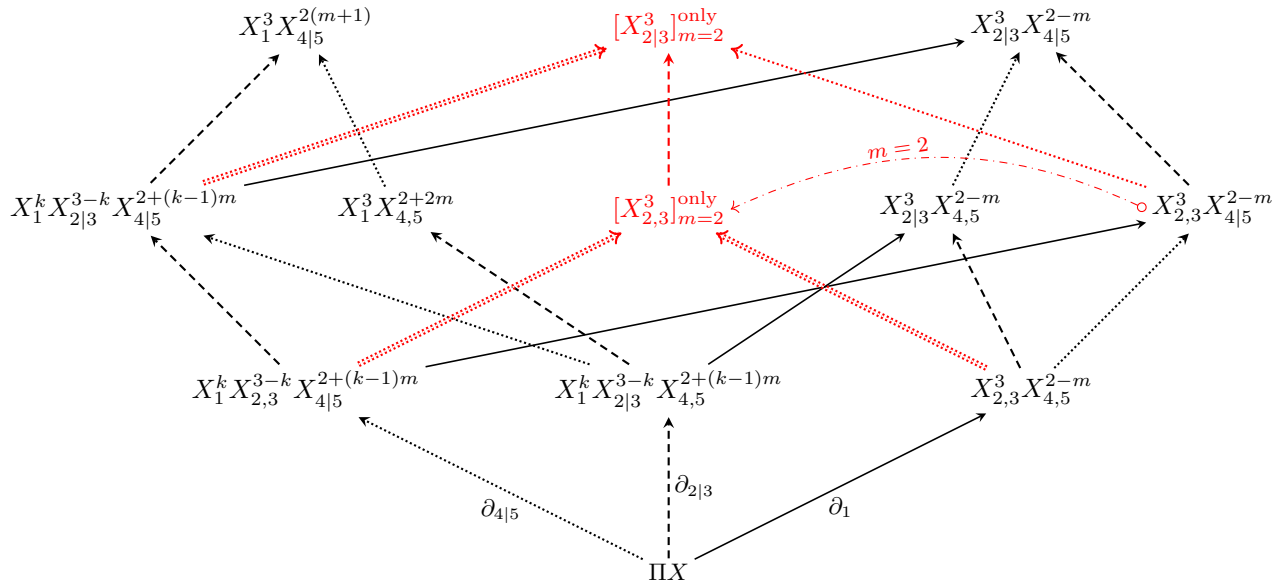

\begin{footnotesize}
\begin{center}
 \TikZ{[thick, every node/.style={inner sep=0,outer sep=.75mm}, scale=1.2]
  \path[use as bounding box](-7,0)--(6.5,6.5);
  \node(Px) at(0,0) {$\Pi X$};
  \node(c1) at( 4,2) {\small$X_{2,3}^3X_{4,5}^{2-m}$};
  \node(c2) at( 0,2) {\small$X_1^kX_{2|3}^{3-k}X_{4,5}^{2+(k-1)m}$};
  \node(c4) at(-4,2) {\small$X_1^kX_{2,3}^{3-k}X_{4|5}^{2+(k-1)m}$};
  \node(c14) at( 6,4) {\small$X_{2,3}^3X_{4|5}^{2-m}$};
  \node(c12) at( 3,4) {\small$X_{2|3}^3X_{4,5}^{2-m}$};
  \node[red](c45) at(0,4) {\small$[X_{2,3}^3]_{m=2}^{\text{only}}$};
  \node(c23) at(-3,4) {\small$X_1^3X_{4,5}^{2+2m}$};
  \node(c24) at(-6,4) {\small$X_1^kX_{2|3}^{3-k}X_{4|5}^{2+(k-1)m}$};
  \node(c124) at( 4,6) {\small$X_{2|3}^3X_{4|5}^{2-m}$};
  \node(c234) at(-4,6) {\small$X_1^3X_{4|5}^{2(m+1)}$};
  \node[red](c345) at(0,6) {\small$[X_{2|3}^3]_{m=2}^{\text{only}}$};
  \draw[semithick, -stealth](Px)--node[below=2pt]
    {\fnSz$\vd_1$}(c1);
  \draw[densely dashed, -stealth](Px)--node[right=0pt]
    {\fnSz$\vd_{2|3}$}(c2);
  \draw[densely dotted, -stealth](Px)--node[below=2pt]
    {\fnSz$\vd_{4|5}$}(c4);
  \draw[semithick, -stealth](c2)--(c12);
  \draw[semithick, -stealth](c4)--(c14);
  \draw[densely dashed, -stealth](c1)--(c12);
  \draw[densely dashed, -stealth](c2)--(c23);
  \draw[densely dashed, -stealth](c4)--(c24);
  \draw[densely dotted, -stealth](c1)--(c14);
  \draw[densely dotted, -stealth](c2)--(c24);
  \draw[red, densely dotted, double, -{>[length=4pt, width=6pt]}](c4)--(c45);
  \draw[red, densely dotted, double, -{>[length=4pt, width=6pt]}](c1)--(c45);
  \draw[semithick, -stealth](c24)--(c124);
  \draw[densely dashed, -stealth](c24)--(c234);
  \draw[red, densely dashed, -stealth](c45)--(c345);
  \draw[densely dashed, -stealth](c14)--(c124);
  \draw[densely dotted, -stealth](c23)--(c234);
  \draw[red, densely dotted, double, -{>[length=4pt, width=6pt]}](c24)--(c345);
  \draw[red, densely dotted, -{>[length=4pt, width=6pt]}](c14)--(c345);
  \draw[densely dotted, -stealth](c12)--(c124);
  \draw[red, thin, dash dot, {Circle[open]}->](c14.west)to[out=155,in=25]
      node[above left=1pt, rotate=8]{\scriptsize$m=2$}(c45.east);
 }
\end{center}
\end{footnotesize}
 \caption{The poset structure of the collections $(\delta_i\Pi X)$, simplified by compressing $(X_2{\oplus}X_3)\mapsto X_{2,3}$, writing $X_{2|3}$ for ``either $X_2$ or $X_3$,'' etc.; for exceptional $[X_{2,3}^3]_{m=2}^{\text{only}}$ and  
$[X_{2|3}^3]_{m=2}^{\text{only}}$, see text}
 \label{f:poset}
\end{figure}

It therefore behooves us to consider the possible deformations of $\Pi X$, which may be generated systematically by
({\bf1})~omitting one $X_i$ from the product $\Pi X$, and
({\bf2})~replacing it with a monomial, $\frm_i(X)$, formed using
$\{X_j{:}~ j\neq i\}$, and which typically includes numerous continuously variable parameters. 
The exclusion of $X_i$ from the monomials $\frm_i(X)$ that replace $X_i$ is directly motivated by analogy with the requirement in quantum mechanics, where perturbative changes of any state, $|*\rangle$, must be orthogonal to 
$|*\rangle$ so as to preserve the norm and unitarity.
The operators $\delta_i:=\frm_i(X)\vd_i$ are then in fact generators of $U(1;\IC)^2$-equivariant coordinate reparametrizations, also cited as (Demazure) ``automorphisms stemming from roots'' (here, the $X_i$)\footnote{Each such automorphism is generated by a relation of the form 
$X_i\!-\!\lambda\frm_i(X)\sim0$, effectively substituting $X_i$ by a multiple of $\frm_i(X)$---which is {\em\/exactly\/} how the $\frm_i(X)\vd_i$ act.}\cite[p.48]{rCK}.
Given the simple form of the starting point, $\Pi X$, the following observations follow:
\begin{enumerate}[topsep=-1pt]\raggedright
 \item 
   The 1st order deformations are all of the form
   $\delta_i\Pi X=\frm_i(X)\,(\vd_i\Pi X)$, with no summation on $i$.
 \item 
   Since both $(\vd_i\Pi X)$ and $\frm_i(X)$ are $X_i$-independent,\\
   so is each monomial in the collection $\delta_i\Pi X$.
 \item \label{i:abut}
   Two distinct collections, $\delta_i\Pi X$ and  $\delta_j\Pi X$, 
   {\em\/adjoin\/} along their common subset of monomials, which then are 
   independent of {\em\/both\/} $X_i$ and $X_j$, and can be reached by both
   $\delta_i\Pi X$- and $\delta_j\Pi X$-deformation paths.
 \item 
   Three distinct collections have in common their subset of monomials 
   that are independent of three distinct $X_i$'s, etc.
 \item \label{i:dd=0}
   Since $\vd_i^2\Pi X\equiv0$, there are no 2nd order deformations to consider.
 \item 
   Combining observations~\ref{i:abut} and~\ref{i:dd=0}, it follows that
   each collection, $\delta_i\Pi X$, is delimited (bounded) by the monomials it 
   shares with precisely one other collection: It retains the 
   $\delta_i\Pi X$-monomials on the side of those common with $\delta_j\Pi X$ 
   that are no farther from $\Pi X$ in the $\vd_j$-direction than 
   $\delta_j\Pi X$ itself.
   Let $[\delta_i\Pi X]$ denote the $i^\text{th}$ so-delimited collection.
\end{enumerate}
The union of these collections with the fundamental monomial,\footnote{By slight abuse of notation, ``$\uplus$'' is herein used to indicate that the overall union retains the complete subset-generated poset structure, such as depicted in Figure~\ref{f:poset}.}
\begin{equation}
  (\Pi X) ~\uplus~ \textstyle\biguplus_i [\delta_i\Pi X]
 \label{e:allK*}
\end{equation}
has, astoundingly, an overall combinatorial structure that is well-nigh identical to the zero locus of $\Pi X$ itself, the polyhoron mentioned above!

The deformation-organizing structure outlined in the above six points is in fact perfectly

\paragraph{Poset and Multitope.}
In fact, the total collection~\eqref{e:allK*} has the structure of a poset, depicted in Figure~\ref{f:poset}, simplified for clarity by denoting the linear combination $X_2\!\oplus\!X_3\mapsto X_{2,3}$, as they have the same charges; similarly for $X_4\!\oplus\!X_5\mapsto X_{4,5}$. 
In turn, $X_{2|3}$ stands for ``either $X_2$ or $X_3$''
and $X_{4|5}$ for ``either $X_4$ or $X_5$.''
For $\FF[3]m$, the 5-variable anticanonical monomials are restricted by the $q_1$- and $q_2$-cancellation conditions~\eqref{e:CY}, and so occupy a portion of a 3-dimensional lattice. In this lattice of monomials, 
each $X_i$-independent subset, $[\delta_i\Pi X]$, forms a ``pane,'' 
each $X_i,X_j$-independent subset forms a ``line'' where two ``panes'' adjoin,
and 
each $X_i,X_j,X_k$-independent subset is a corner where those three ``panes'' adjoin.
Since ``panes'' of monomials are delimited by the ``lines'' of monomials that are shared with another ``pane,'' non-lattice (``fractional'') intersections of lattice planes containing the monomial ``panes'' do not in fact delimit any of the combinatorial strata of the poset.

Already from the $q_a$-charges~\eqref{e:3F5}, it is immediate that there can exist no $X_1,X_2,X_3$-independent deg-$\pM{3\\2-m}$ monomials---which in fact reflects a generator of the so-called Stanley-Reisner ideal\cite{rCLS-TV}. In turn, rendering any of the monomials in Figure~\ref{f:poset} $X_4,X_5$-independent can only happen for the exceptional value, $m=2$: Reading from the left in the 2nd layer from below:
\begin{equation}
  X_1^kX_{2,3}^{3-k}X_{4|5}^{2+(k-1)m}
  \xrightarrow{2+(k-1)m=0}~
  X_1^{1-2/m}X_{2,3}^{2+2/m},
 \label{e:m=1,2}
\end{equation}
which can have non-fractional exponents only for $m=1,2$. The first of these, however, results in $X_1^{-1}X_{2,3}^4$, which is unacceptable: via~\eqref{e:2P-T}, $X_1=\frs_0(x,y)$ in the biprojective realization~\eqref{e:pe(x,y)}, and only non-negative (integral) powers of $\frs_0(x,y)$ are holomorphic; see Remark~\ref{r:no1/dx}. This leaves $m=2$ as the only choice in~\eqref{e:m=1,2}---which then also becomes $X_1$-independent. Rendering the middle ``pane'' of monomials in Figure~\ref{f:poset} $X_4,X_5$-independent leads back to~\eqref{e:m=1,2}, and we are left with the rightmost ``pane,'' where the inference is immediate:
\begin{equation}
  X_{2,3}^3X_{4,5}^{2-m}
  \xrightarrow{2-m=0}~
 [X_{2,3}^3]_{m=2}^{\text{only}}.
 \label{e:m=2only}
\end{equation}
The automatic loss of dependence on two $X_i$'s is depicted in Figure~\ref{f:poset} by doubled arrows. Finally, the $m=2$-labeled arcing arrow merely denotes that $[X_{2,3}^3]_{m=2}^{\text{only}}$ is the indicated special case of $X_{2,3}^3X_{4|5}^{m-2}$.

The poset shown in Figure~\ref{f:poset} details the combinatorial structure of the collection of monomials that form the so-called {\em\/Newton multitope.}\footnote{\label{fn:mtp}{\em Multitope\/} 
is the heterotic portmanteau contraction of ``(possibly)
\underline{multi}-layered poly\underline{\smash{tope}} complex,'' which may be
self-crossing, i.e., flip-folding and/or otherwise multi-layered and may be 
wrapping the origin more (or fewer) than once. The well-known (complete)
{\em\/polytopes\/} are {\em\/plain\/} (single-layered) special cases of
multitopes wrapping the origin precisely once.} For (complex-algebraic) toric
varieties, this monomial collection forms the {\em\/standard\/} Newton polytope,
which is always a convex polygon; in particular, this standard part for the case
at hand and for $m\geqslant3$ {\em\/omits\/} the the (now {\em\/rational!\/}) right-hand side monomials,
\begin{equation}
  X_{2,3}^3X_{4,5}^{2-m},\qquad
  X_{2|3}^3X_{4,5}^{2-m},~~ X_{2,3}^3X_{4|5}^{2-m},\qquad
   X_{2|3}^3X_{4|5}^{2-m}.
 \label{e:rK*}
\end{equation}
This omission exposes the central $\Pi X$ into a facet of this standard Newton polytope, which in turn implies that no so-constructed polynomial is transverse and all {\em\/regular\/} Calabi--Yau hypersurfaces must be singular.

\begin{figure}[htb]
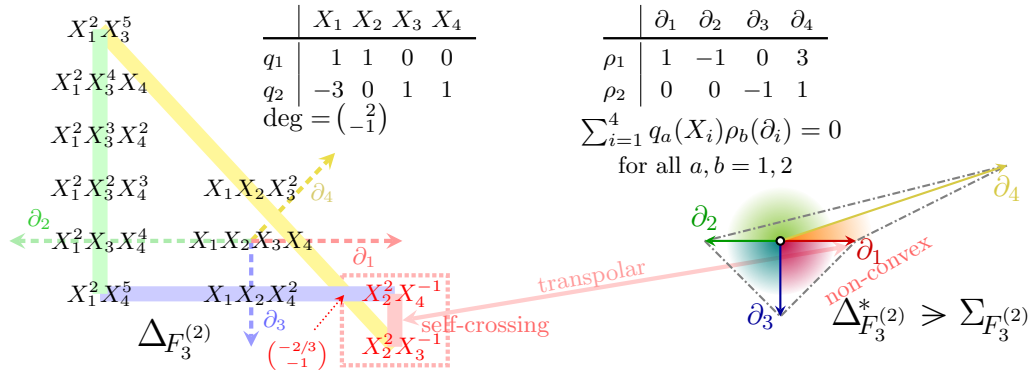

\centering
\TikZ{[ultra thick]
      \path[use as bounding box](-3.2,-1.5)--(10.1,3);
      \begin{scope}[yscale=.7]
      \draw[red!30, densely dotted](1.2,-.65)rectangle++(1.4,-1.7);
      \path[red!60](3.1,-1.6)node{\footnotesize self-crossing};
      \draw[yellow!80!gray, densely dashed, -stealth](0,0)--({atan(3/2)}:2);
       \path[yellow!60!gray]({atan(3/2)}:1.7)node[below=2pt]
       {\footnotesize$\vd_4$};
      \draw[blue!40, densely dashed, -stealth](0,0)--(-90:2);
       \path[blue!60](-90:1.5)node[right]{\footnotesize$\vd_3$};
      \draw[red!40, densely dashed, -stealth](0,0)--(0:2);
       \path[red!60](0:1.5)node[below=-2pt]{\footnotesize$\vd_1$};
      \draw[green!60!gray!40, densely dashed, -stealth](0,0)--(180:3.2);
       \path[green!60!gray](180:2.8)node[above]{\footnotesize$\vd_2$};
      \draw[yellow!50, line width=2mm](-2,4)--(1.9,-2);
      \draw[blue!20, line width=2mm](1.9,-1)--(-2,-1);
      \draw[green!20, line width=2mm](-2,-1)--(-2,4);
      \draw[red!20, line width=2mm](1.9,-1)--(1.9,-2);
      \path(-2,4)node{\footnotesize$X_1^2X_3^5$};
      \path(-2,3)node{\footnotesize$X_1^2X_3^4X_4$};
      \path(-2,2)node{\footnotesize$X_1^2X_3^3X_4^2$};
      \path(-2,1)node{\footnotesize$X_1^2X_3^2X_4^3$};
      \path(-2,0)node{\footnotesize$X_1^2X_3X_4^4$};
      \path(-2,-1)node{\footnotesize$X_1^2X_4^5$};
      \path(0,1)node{\footnotesize$X_1X_2X_3^2$};
      \path(0,0)node{\footnotesize$X_1X_2X_3X_4$};
      \path(0,-1)node{\footnotesize$X_1X_2X_4^2$};
      \path[red](2,-1)node{\footnotesize$X_2^2X_4^{-1}$};
      \path[red](2,-2)node{\footnotesize$X_2^2X_3^{-1}$};
      \path(0,2.3)node[right]{\small$\deg=\pM{~~2\\-1}$};
      \path(-1,-1.8)node{\Large$\pDN{\FF[2]3}$};
      \path(0,3.5)node[right]
      {\footnotesize$\begin{array}[t]{@{}r|c@{~}c@{~}c@{~}c}
                         &X_1 &X_2 &X_3 &X_4\\ \toprule\noalign{\vglue-2pt}
                     q_1 &~~1 & 1  & 0  & 0 \\
                     q_2 &-3  & 0  & 1  & 1 \\
                       \end{array}$};
      \path(4.5,3.5)node[right]
      {\footnotesize$\begin{array}[t]{@{}r|c@{~~}c@{~~}c@{~~}c}
                  &\vd_1 &\vd_2 &\vd_3 &\vd_4\\ \toprule\noalign{\vglue-2pt}
           \rho_1 & 1 & -1 &~0 & 3 \\
           \rho_2 & 0 &~0 & -1 & 1 \\
                     \end{array}$};
      \path(4.2,2.1)node[right]{\small$\sum_{i=1}^4q_a(X_i)\rho_b(\vd_i)=0$};
      \path(4.7,1.4)node[right]{\footnotesize for all $a,b=1,2$};
      \draw[red!20, stealth-stealth](2,-1.5)--
        node[above=-6pt, xshift=-11pt, rotate=10, red!40]
        {\footnotesize transpolar}++(5.9,1.4);
      \draw[red, semithick,  densely dotted, stealth-](1.23,-1.02)--++(-.4,-.67)
        node[below, xshift=-2mm]{\tiny$\pM{-\sfrac23\\-1}$};
    \end{scope}
    \begin{scope}[xshift=7cm, thick]
      \corner{(0,0)}{0}{atan(1/3)}{1.2}{orange};
      \corner{(0,0)}{atan(1/3)}{180}{.7}{yellow!50!green};
      \corner{(0,0)}{180}{270}{.7}{teal};
      \corner{(0,0)}{270}{360}{.7}{purple};
      \draw[gray, densely dash dot, line join=round]
        (-1,0)--(0,-1)--(1,0)--(3,1)--cycle;
      \draw[red!80!black, -stealth](0,0)--(1,0)
        node[right=-4pt, yshift=-4pt]{$\vd_1$};
      \draw[green!60!black, -stealth](0,0)--(-1,0)
        node[above]{$\vd_2$};
      \draw[blue!60!black, -stealth](0,0)--(0,-1)
        node[left=-2pt]{$\vd_3$};
      \draw[yellow!80!black, -stealth](0,0)--(3,1)
        node[below]{$\vd_4$};
      \filldraw[fill=white](0,0)circle(.5mm);
      \path(2,-1)node{\Large$\pDs{\FF[2]3}\mathrel{-\mkern-19mu>}\pFn{\FF[2]3}$};
      \path[red!60](1.3,-.4)node[rotate=30]{\footnotesize non-convex};
    \end{scope}
}
 \caption{First-order deformations, $[\delta_i\Pi X]$,
          shown here for $\FF[2]3$, compressed vertically for space.
          The $\vd_i$-directions are orthogonal to $[\delta_i\Pi X]$,
          and re-drawn at right, at proper relative scale and angle.}
 \label{f:2F3K*}
\end{figure}

\paragraph{In Two Dimensions}
These suggestive characteristics are easier to plot and see for the 2-dimensional case of Hirzebruch surface scrolls, $\FF[2]m$; see  Figure~\ref{f:2F3K*}, where the simplest nontrivial $m=3$ case is shown, to save space.
Indeed, this rationale precisely recovers all regular monomials but also the specific rational monomials associated with the (red-ink indicated) ``extension'' of this (Newton) multitope\cite{rBH-gB, Berglund:2022dgb, Berglund:2024zuz, Hubsch:2025sph, Hubsch:2025teh, Hubsch:2026wxe}. A baker's dozen of straightforward observations are then in order:
\begin{enumerate}[topsep=-1pt]

\item 
 The collection of monomials, 
 $[\delta_4\Pi X]=\{X_1^2X_3^5,X_1X_2X_3^2,X_2^2X_3^{-1}\}$,
 along the slanted (yellow-ink highlighted) ``line'' is
 on the right-hand side delimited by $X_2^2X_3^{-1}$, since this is
 common with the right-hand side (red-ink indicated) vertical collection,
 $[\delta_1\Pi X]=\{X_2^2X_4^{-1},X_2^2X_3^{-1}\}$.
 Next in $\delta_4\Pi X$, $X_1^{-1}X_2^3X_3^{-4}$ appears to the right of
 the already (rightmost) $\vd_1$-distance-1 (vertical, pink-highlighted) ``line'' 
 $\delta_1\Pi X$, and so  is``beyond'' the juncture, 
 $X_2^2X_3^{-1}=[\delta_4\Pi X]\sqcap[\delta_1\Pi X]$.

\item 
 The collection of monomials, 
 $[\delta_3\Pi X]=\{X_1^2X_4^5,X_1X_2X_4^2,X_2^2X_4^{-1}\}$,
 along the horizontal (blue-ink indicated) ``line'' is
 on the right-hand side delimited by $X_2^2X_4^{-1}$, since this is
 common with the right-hand side (red-ink indicated) vertical collection,
 $[\delta_1\Pi X]=\{X_2^2X_4^{-1},X_2^2X_3^{-1}\}$.
 Next in $\delta_3\Pi X$, $X_1^{-1}X_2^3X_4^{-4}$ similarly appears to the right 
 of $\delta_1\Pi X$, and so  is``beyond'' the juncture, 
 $X_2^2X_4^{-1}=[\delta_3\Pi X]\sqcap[\delta_1\Pi X]$.

\item 
 The non-lattice ``intersection,'' $(\sfrac{2\!}3,{-}1)$, of the
 slanted (yellow) and horizontal (blue) lines does not indicate a common 
 monomial and so does not delimit any collection of monomials.
 Related to this, the fact that 
 $X_2^2X_3^{-1}=[\delta_4\Pi X]\sqcap[\delta_1\Pi X]$ is below $\delta_3\Pi X$
 does not preclude deforming to $X_2^2X_3^{-1}$ via these two adjoining 
 deformation ``lines,'' which are independent of $\delta_3\Pi X$. 
 Similarly, the fact that
 $X_2^2X_4^{-1}=[\delta_3\Pi X]\sqcap[\delta_1\Pi X]$ is above $\delta_4\Pi X$
 does not preclude deforming to $X_2^2X_4^{-1}$ via these two adjoining 
 deformation ``lines,'' which are independent of $\delta_4\Pi X$.

\item \label{i:d=1}
 Consisting of 1st order deformations of $\Pi X$, each ``line,''
 $[\delta_i\Pi X]$, is at unit distance from $\Pi X$, as is each facet of every
 reflexive polytope\cite{rBaty01}.

\item 
 The ``lines'' $[\delta_i\Pi X]$ form a (self-crossing on the right-hand 
 side\footnote{This is a general property for all $\FF{m}$ with $m\ge3$.}) 
 polygonal loop, $\pDN{\FF[2]3}$, enclosing $\Pi X$; such loops are called 
 {\em\/generalized legal loops\/}\cite{rP+RV-12}, and are the 2-dimensional 
 case of {\em\/VEX multitopes\/}\cite{rBH-gB, Berglund:2022dgb, 
 Berglund:2024zuz, Hubsch:2025sph, Hubsch:2025teh, Hubsch:2026wxe}.
 Jointly with the previous, unit-distance observation~\ref{i:d=1}, this implies
 that VEX multitoeps (see footnote~\ref{fn:mtp}) directly generalize reflexive 
 polytopes.

\item 
 The evident (e.g.,\ \textsc{ccw}) polygonal loop of ``lines'' 
 $[\delta_i\Pi X]$ specifies the orientation-defining order, $\{1,4,2,3\}$,
 which induces the \textsc{ccw}-orientation and ordering in
 $\{\vd_1,\vd_4,\vd_2,\vd_3\}$---which then also form a polygonal loop,
 $\pDs{\FF[2]3}$, with edges at unit distance from the origin.

\item 
 $\pDN{\scX}$ is the {\em\/Newton multitope,} and $\pDs{\scX}$ the fan-spanning
 multitope of $\scX$: it spans the ($\vd_i$-)fan $\pFn{\scX}$, as
 depicted in Figure~\ref{f:2F3K*} (bottom right). If $\pFn{\scX}$ is
 {\em\/plain\/} (single-layered), it uniquely specifies the underlying toric
 variety\cite{rD-TV, rF-TV}; the multi-layered cases will be discussed below.

\item 
 Each deformation direction, $\vd_i$, acts along a lattice-vector, 
 $\skew0\vec\rho(\vd_i)$, its components $(\rho_1,\rho_2)$ given in 
 Figure~\ref{f:2F3K*} (top right), which is orthogonal to
 its ``line,'' $[\delta_i\Pi X]$. They span the $\vd_i$-fan depicted in
 Figure~\ref{f:2F3K*}, bottom right.

\item 
 The coordinate field charges and the lattice components of the 
 corresponding directions are mutually orthogonal:
 $\sum_{i=1}^4 q_a(X_i)\rho_b(\vd_i)=0$: The pair of $(n{+}2)$-vectors 
 $\rho_b$ spans the null-space of the pair $q_a$, and vice versa.

\item 
 the 2-vectors 
 $\big\{\big(\rho_1(\vd_i),\rho_2(\vd_i)\big),~i\!=\!1,\!\cdots\!,5\big\}$ 
 tabulated in Figure~\ref{f:2F3K*} (top right) precisely reconstruct (up to a
 $180^\circ$ rotation) the corresponding vectors $\nu_i$ as shown in the 
 $n\!=\!2$ version of~\eqref{e:q1-4}.

\item 
 By construction leading to the $\vd_i$-fan and enveloping (``spanning'' 
 polygon, $\pDs{\FF[2]3}$, in Figure~\ref{f:2F3K*} (bottom right), the vertices
 of any so-constructed $\pDs{\scX}$ are in 1--1 correspondence with the GLSM
 chiral superfields, $X_i$---which then are identified as the Cox coordinates 
 of $\scX$.

\item 
 The fact that the $q_a(X_i)$-charges preclude the existence of
$X_1,X_2,X_3$- and $X_4,X_5$-independent deg-$\pM{3\\2-m}$ monomials (as discussed above and except for the $m=2$ case~\eqref{e:m=2only}) reproduces the generators of the so-called Stanley-Reisner ideal of $\FF[3]{m}$\cite{rCLS-TV}.

\item 
 $\pDN{\scX}$ and $\pDs{\scX}$ are the {\em\/transpolar\/} of one 
 another\cite{rBH-gB} and\cite[Def.\:1.2]{Berglund:2024zuz}.
\end{enumerate}

\section{Conclusions, Outlook, Discussion and Aspirations}
\label{s:CODA}
The foregoing presentation details the construction of worldsheet QFT models, and especially the GLSM type, the ground states of which are Ricci-flat/Calabi--Yau manifolds found as the zero locus of suitably chosen superpotentials. This embeds them in suitably well known ``ambient spaces,'' in turn specified by the same worldsheet QFT model with the superpotential omitted; see Remark~\ref{r:noW}.
This retrospective analysis indicates room for extension and motivates the generalization beyond the complex-algebraic toric geometry that has so successfully served for the past four decades. With some of these generalizations indicated in Remarks~\ref{r:nFlp} and~\ref{r:nTV}, we now turn to the universal motivation\cite{rBH-gB, Berglund:2022dgb, Berglund:2024zuz, Hubsch:2025sph, Hubsch:2025teh, Hubsch:2026wxe}, mirror symmetry.

\subsection{Into the Mirror\dots{}}
The ``Newton multitope,'' of 1st-order deformations of the fundamental monomial, $\Pi X$, and plotted on the left-hand side of Figure~\ref{f:2F3K*}, is self-crossing also in higher dimensions, $n\geq2$, and so for all $\FF{m}$ with $m\geq3$.
The standard ``polar'' operation of toric geometry then must be adapted to the (face-wise iterative) ``transpolar'' operation\cite{rBH-gB, Berglund:2022dgb, Berglund:2024zuz} to reconstruct the (well known) fan of 
$\FF[2]3$---reproducing the non-convex region adjacent to the concave vertex, 
$\rho(\vd_1)$, and which then perfectly agrees with the $\vd_i$-fan depicted in~\eqref{f:2F3K*}, lower right.
For all complex-algebraic toric varieties, that $\vd_i$-fan is indeed always {\em\/plain\/} (it is neither self-crossing, flip-folding nor otherwise multi-layered), and fully specifies the underlying toric variety on which the corresponding collection of monomials serve as anticanonical sections. The polytope bounded by the bases of the cones generated by consecutive pairs of vectors, $[\rho(\vd_i),\rho(\vd_{i+1})]$, and dot-and-dash gray-outlined in Figure~\ref{f:2F3K*} is said to span the $\vd_i$-fan and is itself plain.

On the other hand, each self-crossing or otherwise multi-layered (but VEX\cite{rBH-gB, Berglund:2022dgb, Berglund:2024zuz}) multitope, $\pDN{\scX}$, such as the Newton multitope on the left-hand side of Figure~\ref{f:2F3K*} analogously spans
a collection of top-dimensional cones, one over each delimited ``pane,'' 
$[\delta_i\Pi X]$, which inherits
the poset structure such as depicted in Figure~\ref{f:poset}
as well as the self-crossing multi-layered features of $\pDN{\scX}$. 
Such $n$-dimensional cone-collections are called {\em\/multifans\/} and correspond to real $2n$-dimensional spaces with ``half-dimensional''
$(S^1)^n$-action ($\to$\,non-gauged $U(1)^n$ in a GLSM) that are {\em\/continuously\/} more general than complex-algebraic varieties\cite{rM-MFans, Masuda:2000aa, rHM-MFs, Civan:2003aa, Masuda:2006aa, rHM-EG+MF, rH-EG+MFs2, Nishimura:2006vs, Ishida:2013aa, Ishida:2013ab, buchstaber2014toric}.
These, so-called {\em\/unitary torus manifolds\/} (UTMs) are thus indicated as the general type of spaces corresponding to the Newton multitope of a given GLSM, amongst which the ambient space of the mirror-GLSM should be found.

The motivation for this contention stems from the straightforward generalization of Batyirev's complex-algebraic toric geometry framework for constructing mirror models, $(Z_{\sss\text{CY}},\chZ_{\sss\text{CY}})$, where the mirror operation is essentially encoded in swapping the roles of 
the Newton polytope, $\pDN{\scX}$, and 
the fan-spanning polytope, $\pDs{\scX}$:
\begin{equation}
\vC{\TikZ{[every node/.style={inner sep=0,outer sep=.75mm}]
    \path[use as bounding box](-5.2,-.8)--(5.2,.7);
    \path(0,0)node{$\displaystyle
     Z_{\sss\text{CY}}\!=\!\{f(x){=}0\}\subset\scX\mapsto~\pDs{\scX}
     \overset\triangledown\longleftrightarrow 
     \pDN{\scX}\,{\overset{\sss\text{def}}=}\,\pDs{\tPX}
     ~\mathop{\reflectbox{$\mapsto$}}\tPX\supset
     \{f^\intercal(y){=}0\}\!=\!\chZ_{\sss\text{CY}}.$};
    \draw[stealth-stealth](-4.7,-.35)--++(0,-.4)--node[above, xshift=-3mm]
    {\scSz transpose-mirror\cite{rBH, rBH-gB, Berglund:2022dgb, Berglund:2024zuz, 
     Hubsch:2025sph, Hubsch:2025teh, Hubsch:2026wxe}}
     ++(8.3,0)--++(0,.4);
    \draw[-stealth](-.3,.25)..controls++(-.5,.6)and++(.3,.6)..(-4.65,.2);
    \draw[-stealth](-1.3,.25)..controls++(.5,.6)and++(-.3,.6)..(3.5,.2);
    }}
 \label{e:MM}
\end{equation}
These correspondences are directly realized by straightforwardly extending the original, 1992 construction of mirror models\cite{rBH} and has been recently detailed directly within the GLSM framework\cite{Hubsch:2026wxe}, where the transpose-mirror ambient space, $\tPX$, is the space corresponding to the transposed GLSM with its (transposed) superpotential omitted; see Remark~\ref{r:noW}.

\subsection{\dots{}and the First Reflections}
The technical difficulty---and open challenge---is to identify, 
for a Calabi--Yau hypersurface in a toric space, 
$Z_{\sss\text{CY}}\!=\!\{f(x){=}0\}\subset\scX$,
the particular (unitary torus) manifold, $\tPX$,
in which the transpose-mirror Calabi--Yau hypersurface is embedded,
$\chZ_{\sss\text{CY}}\!=\!\{f^\intercal(y){=}0\}\subset\tPX$.
To this end, the (unitary torus) manifold $\tPX$ should correspond to the (multi)fan spanned by $\pDs{\tPX}:=\pDN{\scX}=(\pDs{\scX})^\triangledown$, as in~\eqref{e:MM}.
When $\pDs{\scX}$ is a (plain and convex) reflexive polytope, so is $\pDN{\scX}$ and the fan it spans defines the transpose-ambient (complex-algebraic) toric variety\cite{rBaty01, rCOK}.
When however $\pDs{\scX}$ is {\em\/not convex,} $\pDN{\scX}$ is a self-crossing multitope, its flip-folded ``extension'' regions directly dual (transpolar\cite{rBH-gB, Berglund:2022dgb, Berglund:2024zuz}) to the non-convex regions in $\pDs{\scX}$.
Such a Newton multitope, $\pDN{\scX}$, spans a multifan which however does not encode a specific space with the multifan-prescribed (non-gauged)
$(U(1)=S^1)^n$-action. In fact, there is a {\em\/continuum\/} of candidates for $\tPX$, among the so-called unitary torus manifolds\cite{rM-MFans, Masuda:2000aa, rHM-MFs}.

This laxness is best seen on comparison: Complex-algebraic toric varieties, $V$, are uniquely encoded by their fans, $\pFn{\,V}$, since:
({\bf1})~each top-dimensional cone of $\pFn{\,V}$ encodes a $\IC^n$-like chart with the standard {\em\/toric\/} $(\IC^*)^n$-action,
({\bf2})~the common face of two adjacent cones,
 $\varsigma=(\sigma\cap\sigma')\in\pFn{\,V}$ encodes the gluing of the two charts, $\mathscr{U}_\sigma\uplus\mathscr{U}_{\sigma'}$, and specifies the transition functions to be complex-algebraic. 
This toric $(\IC^*)^n$-action is identified with the ({\em\/non-gauged\/}) $U(1;\IC)^n$-symmetries; see~\eqref{e:qPn}, \eqref{e:q1-4} and the surrounding discussion. In particular, the worldsheet supersymmetry-induced complexification of the gauge symmetry group, $U(1)\to U(1;\IC)$, links the standard phase transformation $X_i\to e^{\ri q_{ai}\vartheta_a}X_i$ with the rigidly paired ``radial'' scaling $X_i\to|\lambda_a|^{q_{ai}}X_i$, jointly recovering the (``radially'' fattened, non-compact) {\em\/toric\/} action,
\begin{equation}
  X_i\to\lambda{\cdot}X_i:=\prod_a\lambda_a^{q_{ai}}X_i,\qquad
  0\neq\lambda_a = |\lambda_a|e^{\ri\vartheta_a}\in\IC^*.
\end{equation}
Unitary {\em\/torus\/} manifolds are defined by specifying only the ``circle''-action (phase transformation) para\-met\-rized by $e^{\ri\vartheta_a}$, leaving the ``radial'' complement arbitrary. Toric varieties are then unitary torus manifolds where each ``radial'' scaling transformation is rigidly tied to a phase transformation.
The arbitrariness of the ``radial'' scaling transformation in unitary torus manifolds then additionally loosens the requirements on the chart-to-chart transition functions from having to be complex-algebraic to being smooth. Such smooth real $2n$-dimensional manifolds with ``half-dimensional'' $(U(1)=S^1)^n$-action are straightforward to model in worldsheet $(0,1)$-supersymmetric\footnote{The worldsheet $(0,1)$-supersymmetry---in fact, supergravity---is needed to universally eliminate the tachyon from the target-spacetime spectrum and associated instabilities, but does not suffice to induce a complex structure in the target spacetime, and so does not induce the complexification of symmetries, $S^1=U(1)\not\to U(1;\IC)=\IC^*$.} QFT models, but are considerably more tedious (and a priori much less amenable to computation) than their much better understood $(0,2)$- and $(2,2)$-supersymmetric counterparts---which is why the latter have hogged the researchers' attention for the past 42 years.\footnote{The computational benefits of the ``rigidity'' of complex-algebraic toric varieties, sprinkled within the continuum of smooth UTMs and even TTMs like rational numbers among the real, reminds of Danilov's characterization of toric varieties as ``frigid toric crystals''~\cite[p.\,100]{rD-TV}.}

The class of ``topological torus manifolds'' (TTMs)\cite{Ishida:2013ab}, and especially their ``nice'' subclass\cite{Yu:2011aa},\footnote{The ``nice'' topological torus manifolds are encoded by a pair, $(\Sigma_b,\Sigma_v)$, where $\Sigma_b$ is a plain fan, $\Sigma_b$ a possibly self-crossing, flip-folded or otherwise multi-layered multifan, and where the 1--1 matching 1-cone generators all satisfy 
$\vec{b}_i\!\equiv\!\vec{v}_i\,\textrm{mod}\,2$; this allows constructing $\Sigma_b$ from $\Sigma_v$, and so effectively construct a ``nice'' TTM to correspond to a UTM multifan. I thank Amin Gholampour and Zengrui Han for alerting me to these manifolds and discussing their properties.} seems to be promising, and is being actively studied: the requisite bundles and their sections are only being classified\cite{Cui:2025Kly, Cui:2025Equ}. 
In this subclass, to each $U(1)=S^1$ phase-transformation encoded by the multifan, $\pFn{\scX}$ (defined within a real lattice, i.e., $\IR^n$ with a $\ZZ^n$-lattice basis), the counterpart ``radial''-action is obtained by extending the multifan to a ``topological fan,'' defined over a ground field, $\mathscr{R}\approx\IC{\times}\ZZ$, parametrized as
$\big[\begin{smallmatrix}b&0\\c&v\end{smallmatrix}\big]$ with $b{+}ic\in\IC$ and $v\in\ZZ$. Then, the $v$-projection, $\pFn[v]{\scX}$ is the multifan encoding the $(U(1)=S^1)^n$-action, while the complementary projection, $\pFn[b+ic]{\scX}$ encodes the counterpart ``radial''-action, and guarantees a smooth combined
$(\IC^*)^n$-action. 

As a special case of unitary torus manifolds, TTMs also have a complex ``anticanonical'' line bundle, $\mathcal{K}^*_{\!\scX}$, with the 
$\mathrm{T}:=(S^1)^n$-equivariant 1st Chern class, 
$c^{\mathrm{T}}_1(\mathcal{K}^*_{\!\scX})=\sum_i\xi_i$, 
where the  $\xi_i\in H^2_{\mathrm{T}}(\scX,\ZZ)$ generate the Chern class and 1--1 correspond to the multifan generators, $\nu_i$---exactly as in the case of (complex-algebraic) toric varieties\cite{rM-MFans, rHM-MFs}. Nevertheless, and owing to the $b,c\in\IR$ part of the TTM encoding of the $(\IC^*)^n$-action, to each multifan there correspond continuously many a priori distinct TTMs. In turn, the range of $(S^1)^n$- and $(\IC^*)^n$-equivariant equivalences (to know which of them are ``the same'' and in precisely what sense) is an open problem now being actively studied.

From the application side, most standard toric geometry computations of interest in applications to string compactification seem to extend straightforwardly---with but one systematic adaptation: the inclusion of multitope ``orientation,'' which flips as one passes through a self-crossing multitope region of odd codimension\cite{rBH-gB, Berglund:2022dgb, Berglund:2024zuz, Hubsch:2025sph, Hubsch:2025teh, Hubsch:2026wxe}; since ordinary polytopes do not self-cross, their orientation is constant and so is not a distinguishing factor.
The multitope orientation seems to be a very specific type of ``omniorientation''\cite{Buchstaber:2001aa, Masuda:2006aa, Ishida:2013aa}, which however has not been studied previously, so that no identification of our $\tPX$ with a specific type of unitary torus manifolds in the mathematics literature is known.
The ability to produce a wealth of computational data to characterize such 
$\tPX$\cite{Berglund:2024zuz} should then help narrowing down the desired specific subclass of unitary/topological torus manifolds to serve as the proper ambient spaces, $\tPX$ in~\eqref{e:MM}, in which to construct the transposition-mirror, 
$\chZ_{\sss\text{CY}}{=}\{f^\intercal(y){=}0\}\subset\tPX$, 
of a given Calabi--Yau hypersurface in toric varieties,
$Z_{\sss\text{CY}}=\{f(x){=}0\}\subset\scX$---and then in fact generalize to both sides of~\eqref{e:MM} being embedded in such unitary/topological torus manifolds.

\smallskip
\paragraph{Acknowledgments.}
First and foremost, I wish to thank Branko Dragovich for inviting me to the $80^{\text{th}}$ birthday celebration of his lifelong dedicated and inspiring research in mathematical physics.
This ongoing project could not have been possible without decades-long collaborations with P.~Berglund and helpful discussions on these topics with Y.~Cui, C.F.~Doran, A.~Gholampour, Z.~Han, K.~Iga, A.~Malmendier. I also thank M.~Masuda for his generous help with torus manifolds and the rich and diverse literature on this topic.
I am grateful to the Department of Mathematics, University of Maryland, College Park, MD, USA, 
and the Physics Department of the Faculty of Natural Sciences of the University of Novi Sad, Serbia, for the recurring hospitality and resources.

\begingroup
\small\frenchspacing\raggedright\baselineskip=13pt plus 1pt minus 1pt
\def\rasp{\leavevmode\raise.45ex\hbox{$\rhook$}}
\providecommand{\href}[2]{#2}\begingroup\raggedright\endgroup

\endgroup

\end{document}